\documentclass[aps,twocolumn,prd,a4paper,superscriptaddress,floatfix]{revtex4-1}
\usepackage{graphicx}
\usepackage{natbib}
\usepackage{subcaption}
\usepackage{float}

\begin{document}

\title{Effects of biases in domain wall network evolution. II. Quantitative analysis}
\author{J. R. C. C. C. Correia}
\email{Jose.Correia@astro.up.pt}
\affiliation{Centro de Astrof\'{\i}sica, Universidade do Porto, Rua das Estrelas, 4150-762 Porto, Portugal}
\affiliation{Instituto de Astrof\'{\i}sica e Ci\^encias do Espa\c co, CAUP, Rua das Estrelas, 4150-762 Porto, Portugal}
\affiliation{Faculdade de Ci\^encias, Universidade do Porto, Rua do Campo Alegre 687, 4169-007 Porto, Portugal}
\author{I. S. C. R. Leite}
\email{Ines.Leite@astro.up.pt}
\affiliation{Centro de Astrof\'{\i}sica, Universidade do Porto, Rua das Estrelas, 4150-762 Porto, Portugal}
\affiliation{Faculdade de Ci\^encias, Universidade do Porto, Rua do Campo Alegre 687, 4169-007 Porto, Portugal}
\author{C. J. A. P. Martins}
\email{Carlos.Martins@astro.up.pt}
\affiliation{Centro de Astrof\'{\i}sica, Universidade do Porto, Rua das Estrelas, 4150-762 Porto, Portugal}
\affiliation{Instituto de Astrof\'{\i}sica e Ci\^encias do Espa\c co, CAUP, Rua das Estrelas, 4150-762 Porto, Portugal}

\date{15 March 2018}

\begin{abstract}
Domain walls form at phase transitions which break discrete symmetries. In a cosmological context they often overclose the universe (contrary to observational evidence), although one may prevent this by introducing biases or forcing anisotropic evolution of the walls. In a previous work [Correia {\it et al.}, Phys.Rev.D90, 023521 (2014)] we numerically studied the evolution of various types of biased domain wall networks in the early universe, confirming that anisotropic networks ultimately reach scaling while those with a biased potential or biased initial conditions decay. We also found that the analytic decay law obtained by Hindmarsh was in good agreement with simulations of biased potentials, but not of biased initial conditions, and suggested that the difference was related to the Gaussian approximation underlying the analytic law. Here we extend our previous work in several ways. For the cases of biased potential and biased initial conditions we study in detail the field distributions in the simulations, confirming that the validity (or not) of the Gaussian approximation is the key difference between the two cases. For anisotropic walls we carry out a more extensive set of numerical simulations and compare them to the canonical velocity-dependent one-scale model for domain walls, finding that the model accurately predicts the linear scaling regime after isotropization. Overall, our analysis provides a quantitative description of the cosmological evolution of these networks.
\end{abstract}
\maketitle

\section{\label{intr}Introduction}

The breaking of spontaneous symmetries at phase transitions that are thought to have happened in the early Universe led to the formation of topological defects \cite{KIBBLE,VSH}. A domain wall is a type of topological defect that occurs whenever a discrete symmetry is spontaneously broken. The study of domain walls has been comparatively neglected since it was first remarked that the restoration of spontaneously broken discrete symmetries at high temperatures in the early universe would lead to a catastrophic scenario \cite{ZEL}. In a nutshell, standard domain walls evolve into the well-known linear attractor scaling solution \cite{VSH,VOSbook}; this has been recently confirmed in high-resolution field theory simulations \cite{Rybak1,Rybak2}. This attractor solution in turn implies that the density of the wall network would end up dominating the energy density of the Universe, which directly contradicts observations, for example, of the Cosmic microwave background \cite{Lazanu}.

In order to avoid this scenario, it is necessary to shorten the lifetime of such defect networks. In a previous paper \cite{Previous} (henceforth Paper I) we used the Press-Ryden-Spergel (PRS) algorithm to numerically simulate various types of biased domain wall networks, aiming to quantify whether (and, if so, how) the linear scaling solution breaks down if the standard initial conditions are biased in one of several ways. Specifically we considered the cases of anisotropic walls \cite{Fossils}, biased initial conditions \cite{Coulson,Larsson}, and a biased potential \cite{Gelmini}. In the anisotropic case we found that the networks still reach scaling, although anisotropic imprints will delay the approach to scaling. On the other hand, with a biased potential or biased initial condition the networks will quickly decay. Interestingly, an analytic model for domain wall network decay, previously proposed by Hindmarsh \cite{Hindmarsh}, was found to be in good agreement with the simulations in the case of a biased potential but not in the case of biased initial conditions, and in Paper I we speculated that the difference could be related to the fact that the Hindmarsh model relies on a Gaussian ansatz for the field's probability distribution function.

Here we provide several quantitative extensions to our Paper I analysis. For anisotropic walls we take advantage of recent improvements in the analytic velocity-dependent one-scale (VOS) model of standard domain wall networks \cite{Rybak1,Rybak2}. We carry out a more extensive set of simulations of anisotropic networks---including the first such simulations using a general purpose graphics-processing-unit (GPU) implementation of the PRS algorithm \cite{GPGPU}--- and compare them to the VOS model, finding that the model accurately predicts the approach of these networks to the linear scaling solution, after isotropization. For the cases of biased potential and biased initial conditions we use our numerical simulations for a detailed study of the field distributions, finding strong evidence in support of the hypothesis that the validity (or not) of the Gaussian approximation is indeed the key difference between the two cases and explains why the Hindmarsh decay law only applies to one of them. Our results in this work therefore lead to a more thorough description of the evolution of these networks; examples of such networks in the context of dynamical supersymmetry breaking and moduli stabilisation are discussed in \cite{LLM}.

The plan of the rest of the paper is as follows. In Sect. \ref{walls} we provide a short overview of domain wall network evolution, both in the standard and in the biased cases. For the latter this is mainly a summary of our Paper I results, which we include here with the goal of making the current work self-contained. We then present our new results for biased networks in Sect. \ref{biases}, and for anisotropic networks in Sect. \ref{anisot}. Finally, Sect. \ref{concl} contains some concluding remarks. Throughout this work we use natural units with $c=\hbar=1$.

\section{\label{walls}Domain wall network evolution overview}

We start by a short review of the basic properties of domain walls, as well as the numerical techniques used in field theory simulations of their cosmological evolution. We will then summarize our previous results in Paper I \cite{Previous}, thereby introducing the key concepts and definitions that will be relevant for the rest of the article.

\subsection{\label{standwall}Standard walls and their evolution}

The simplest field theory model with domain wall solutions has a single real scalar field $\phi$ with Lagrangian density
\begin{equation}
\mathcal{L}=\frac{1}{2}(\partial_\mu\phi)(\partial^\mu\phi)-V_0\left({\frac{\phi^{2}}{\phi_{0}^{2}}}-1\right)^{2}\,.
\label{potential}
\end{equation}
The height of the potential barrier, wall surface tension and wall thickness are, respectively
\begin{equation}
V_0=\frac{\lambda}{4}\phi_{0}^4\,,
\end{equation}
\begin{equation}
\sigma=\frac{2\sqrt{2}}{3}\sqrt{\lambda}\phi_{0}^3\,,
\end{equation}
and
\begin{equation}
\delta\sim\frac{\phi_{0}}{\sqrt{V_0}}=\frac{2}{\sqrt{\lambda}\phi_{0}}\,.
\end{equation}

Using standard variational methods one obtains the field equation of motion. In flat homogeneous and isotropic Friedmann-Lemaitre-Robertson-Walker (FLRW) universes this can be written in terms of physical time $t$ as follows
\begin{equation}
{\frac{{\partial^{2}\phi}}{\partial t^{2}}}+3H{\frac{{\partial\phi}}{\partial t}}-\nabla^{2}\phi=-{\frac{{\partial V}}{\partial\phi}}\,.\label{dynamics}
\end{equation}
where $\nabla$ is the Laplacian in physical coordinates, $H=a^{-1}(da/dt)$ is the Hubble parameter and $a$ is the scale factor. In the numerical simulations discussed in what follows, we assume that the scale factor varies as a power law $a \propto t^\beta$; for example, in the radiation era we have $\beta=1/2$, while in the matter era $\beta=2/3$.

Our field theory simulations are based on the PRS algorithm \cite{Press}: the equations of motion are modified in such a way that the thickness of the domain walls is fixed in co-moving coordinates, while ensuring that relevant conserved quantities are preserved. This has the numerical advantage of enabling a larger dynamic range to be simulated---a clear advantage for studying network evolution. One expects that this will have a small impact on the large scale dynamics of the domain walls, since a wall's integrated surface density (and surface tension) are independent of its thickness. In particular, this assumption should not affect the presence or absence of a scaling solution, provided one uses a minimum thickness  \cite{Press,Leite}. Analogous results have been found for cosmic strings \cite{Moore,Daverio}.

Thus in the PRS method, equation (\ref{dynamics}) becomes
\begin{equation}
{\frac{{\partial^{2}\phi}}{\partial\eta^{2}}}+\alpha_1\left(\frac{d\ln
a}{d\ln\eta}\right){\frac{{\partial\phi}}{\partial\eta}}-{\nabla}^{2}\phi=
-a^{\alpha_2}{\frac{{\partial
V}}{\partial\phi}}\,.\label{dynamics2}
\end{equation}
where $\eta$ is the conformal time and the $\alpha_i$ and are constants. The standard evolution would have $\alpha_i=2$, but in the PRS algorithm $\alpha_2=0$ is used in order to have constant comoving thickness, which then requires $\alpha_1=3$ in order to maintain momentum conservation in an expanding universe. Integration is done via a standard finite-difference scheme.

The standard choice of initial conditions assumes $\phi$ to be a random variable between $-\phi_{0}$ and $+\phi_{0}$ and the initial value of $\partial\phi/\partial\eta$ to be zero. These simple initial conditions will be washed away on a timescale proportional to the wall thickness light crossing time. The specific methods used for generating anisotropic and biased initial conditions are described in the next subsections.

In order to allow the production of several runs with different expansion rates for each of the different cases for anisotropic walls (isotropic, super-horizon isotropic and anisotropic, to be described below), we used our recently developed GPGPU network evolution code (see \cite{GPGPU}, where its validation, speed-ups and bottlenecks are described). The simulations ran on a NVIDIA Quadro P5000, with 2560 CUDA cores, a core clock of 1607 MHz and 16384 MB of memory, clocked at 1126 MHz.

The conformal time evolution of the co-moving correlation length of the network, $\xi_c$ (specifically $A/V_{box}\propto \xi_{c}^{-1}$, $A$ being the comoving area of the walls and $V_{box}$ the box volume) and the box-averaged root-mean squared wall velocity (specifically $\gamma v$, where $ \gamma=1/\sqrt{1-v^2}$ is the Lorentz factor) are directly measured from the simulations \cite{Leite}, and provide our diagnostics for scaling of the networks. Specifically, one looks for the best fit to the power laws
\begin{equation}
\frac{A}{V}\propto\rho_{w}\propto\frac{1}{\xi_{c}}\propto\eta^{\mu}\,,
\label{fit1}
\end{equation}
\begin{equation}
\gamma v\propto\eta^{\nu}\,.
\label{fit2}
\end{equation}
For a scale-invariant behavior, which is the attractor solution for a standard domain wall network in an expanding universe with $a \propto t^\beta$ and $0\le\beta<1$, we expect to have $\mu=-1$ and $\nu=0$, though a sufficiently large dynamic range (and, therefore, a sufficiently large simulation box size) is needed for the simulations to reach this regime in a statistically convincing way \cite{Previous,Rybak1,Leite}.

\subsection{\label{oldfossils}Anisotropic walls}

A plausible cosmological scenario for anisotropic domain wall networks is that they are produced during an anisotropic phase in the early universe, and are subsequently pushed outside the horizon (and thus freeze-out in comoving coordinates) due to an inflationary phase. In this case they will temporarily retain the imprints of this anisotropy, which will only be erased once they re-enter the horizon and become relativistic. This scenario was qualitatively discussed in \cite{Fossils} (which was limited by the small size of its simulations), and Paper I provided more quantitative evidence for this isotropization process.

Since our interest is to study the evolution of the networks in the recent (post-inflationary) universe, we do not need to simulate anisotropic universes, but only initially anisotropic networks evolving in an isotropic universe. Numerically we need to compare three cases:
\begin{itemize}
\item Standard networks, generated with initial conditions as described in the previous subsection, henceforth denoted Case A. These provide our fiducial comparison point, to control for box size effects and other possible numerical artifacts.
\item Super-horizon isotropic networks, which start evolving at $\eta_i=1$  with initial conditions obtained from standard simulations at the conformal time $\eta_h=20$ and with velocities reset to zero, henceforth denoted Case B. The choice of $\eta_h=20$ (which is twice the wall thickness) corresponds to a time where the network is reasonably well defined. This setup allows us to study isotropic networks initially outside the horizon which subsequently re-enter.
\item Anisotropic networks, with initial conditions as in case B, but stretched in one spatial direction by a chosen factor $f$. In Paper I the choice $f=2$ was denoted Case C, while $f=16$ was denoted Case D. In the present work Case D will instead correspond to $f=4$.
\end{itemize}

The analysis of Paper I was based on sets of ten $8192^2$ matter era simulations (smaller box sizes were also simulated), and it confirmed that within the statistical uncertainties the networks do reach scaling, with the timescale needed for the convergence to scaling being larger for the more anisotropic networks (that is, those with a large stretch factor $f$). We now extend this analysis by studying whether the standard way to describe network evolution works for these anisotropic networks. The canonical analytic model for the evolution of network averaged quantities is the Velocity-dependent One-Scale model (VOS) \cite{VOSbook}. This quantitative analytic model was first obtained for cosmic strings \cite{MartinsVOS}, later obtained for domain walls from approximate energy conservation arguments \cite{VOSwalls}, and finally derived rigorously ab initio in \cite{LLM2,Rybak1}. To carry out this detailed study we simulate a range of different expansion rates $\beta$; such a range is important for an accurate calibration of the model, as demonstrated in \cite{Rybak2}.

\subsection{\label{skewed} Biases: initial conditions and potential}

The standard choice of initial conditions for the scalar field $\phi$ assumes that it is uniformly distributed between the two minima of the potential, $\pm\phi_{0}$. In this case we have equal initial fractions of the simulation box in either minimum, and these equal fractions are maintained by the subsequent evolution. Numerically, one can trivially bias the initial conditions by changing the above fractions. Specifically, one can generate initial conditions where the field is uniformly distributed between $-\phi_{0}$ and $+b\phi_{0}$. When $b=1$ we recover the standard (unbiased) case discussed above, while when $b=0$ the whole initial box starts on the same side of the potential and we have no domain walls. (This provides a simple but nevertheless useful way to validate the code.) Thus the initial population fractions in the negative and positive minima will be, respectively,
\begin{equation}
f_-=\frac{1}{1+b}\,,\quad f_+=\frac{b}{1+b}\,; \label{popbias}
\end{equation}
equivalently, we can define a bias parameter
\begin{equation}
\epsilon=\frac{1-b}{2(1+b)}\,. \label{ourepsil}
\end{equation}

Physically, a previous inflationary phase could again be responsible for this, by creating Hubble volumes with slightly different occupation fractions. In this biased case, the domain wall network may transiently evolve as in the standard case, but it will eventually disappear. This scenario was first studied in \cite{Coulson,Larsson}, which were again limited by the box size of their simulations---no larger than $1024^2$. Nevertheless, Coulson \textit{et al.} \cite{Coulson} suggest that for a weak bias (corresponding to population fractions close to $50\%$) a good fit is provided by
\begin{equation}
\frac{A}{V}\propto\eta^{-1}\exp{\left(-\eta/\eta_c\right)}\,, \label{coulson1}
\end{equation}
while for a stronger bias
\begin{equation}
\frac{A}{V}\propto\exp{\left(-\eta/\eta_c\right)}\,, \label{coulson2}
\end{equation}
is sufficient; note the linear dependence on conformal time $\eta$ in the exponential term in both cases.

Later on, Hindmarsh \cite{Hindmarsh} provided analytic arguments suggesting that in the weak bias limit one would should have
\begin{equation}
\frac{A}{V}\propto\eta^{-1}\exp{\left[-\kappa(\epsilon\eta)^n\right]}\,; \label{coulson3}
\end{equation}
here $\epsilon$ is the bias parameter defined in Eq. \ref{ourepsil}, while $\kappa$ is a normalization factor. The exponent $n$ is the number of spatial dimensions, which in the case of the simulations being discussed is always $n=2$. This analysis stems from a relativistic generalization of the condensed matter mean-field approximation method of \cite{OJK}, and relies on a Gaussian ansatz for the field's probability distribution function. Note that in this case the dependence on conformal time appearing in the exponential is quadratic rather than linear, so one expects that the two are numerically distinguishable.

Our analysis in Paper I was based on $2048^2$ matter era simulations, thus alleviating the box size problem. We found that the phenomenological formulas of Coulson \textit{et al.} provide very good fits, unlike that of Hindmarsh: in a statistical sense it is clear that the square dependence on conformal time in the exponential fitting formula is incorrect in this case.

Another way of introducing a bias is through an asymmetry between the two minima of the potential \cite{Gelmini,Larsson}. In this case the  volume pressure from the biasing provides an additional mechanism which affects the dynamics of these walls. A simple  tilted potential is
\begin{equation}
V(\phi)=V_0\left[\left(\frac{\phi^2}{\phi_{0}^2}-1\right)^2+\theta\frac{\phi}{\phi_{0}}\right]\,,
\end{equation}
At early times the wall surface tension dominates, and as long as this is the case the network will have the standard behavior. On the other hand, when the domains become large enough the volume pressure from the energy difference between the two minima will dominate and the walls will decay. This happens at a conformal time
\begin{equation}
\eta_{decay}\sim\frac{\phi_{0}}{\theta\sqrt{V_0}}\sim\frac{2.25}{\mu}\,,\label{decaytimes}
\end{equation}
where the last expression applies for our choice of numerical parameters, both in Paper I and in the present work.

This case was first studied by Larsson \textit{et al.} \cite{Larsson}. Using $1024^2$ simulations, values of $\theta$ up to 0.015, and assuming the fitting function
\begin{equation}
\frac{A}{V}\propto\eta^{-1}\exp{\left[-\kappa(\theta\eta){}^m\right]}\,, \label{coulson4}
\end{equation}
they found that a good fit is provided by an exponent $m=2\pm1$. Note that this is consistent with $m=2$, which corresponds to the Hindmarsh fitting formula, cf. Eq. \ref{coulson3}. Our analysis in Paper I, using $2048^2$ matter era simulations and values of $\theta$ up to 0.1, confirmed that the Hindmarsh choice $m=2$ provides good fits (unlike other integer values of $m$) and also allowed us to measure the normalization parameter $\kappa=(6.34\pm0.01)\times10^{-3}$.

Therefore the most interesting outcome of the analysis in Paper I is that the decay mechanisms for networks with biased initial conditions or a biased potential, are seemingly different, and in particular only the latter is well described by the Hindmarsh analytic formula. In Paper I we speculated that this could be related to Gaussian ansatz for the field probability distribution, which is essential for its derivation \cite{Hindmarsh,OJK}. The further analysis that follows confirms that this is indeed the case.

\begin{figure*}
\includegraphics[width=5cm]{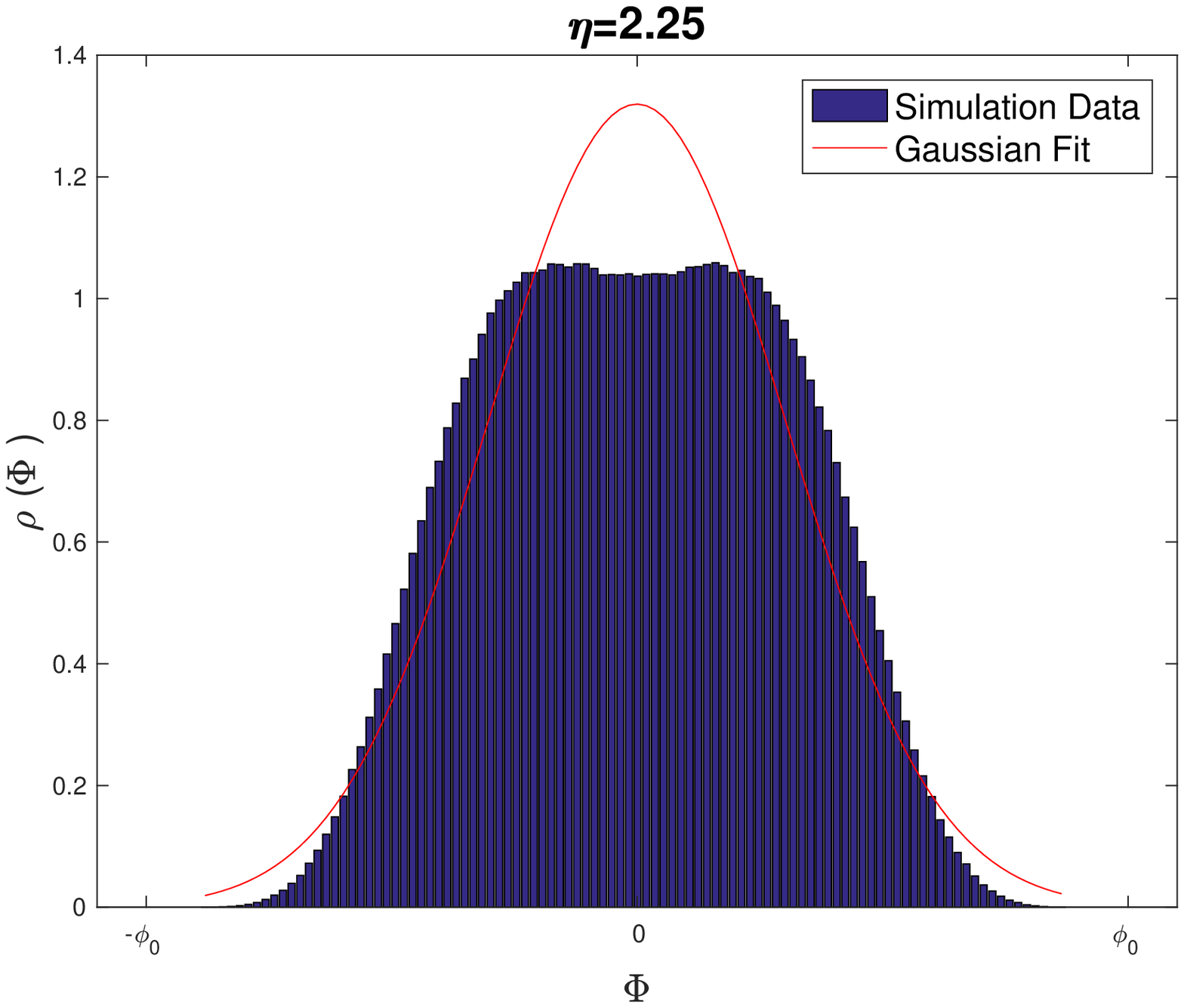}
\includegraphics[width=5cm]{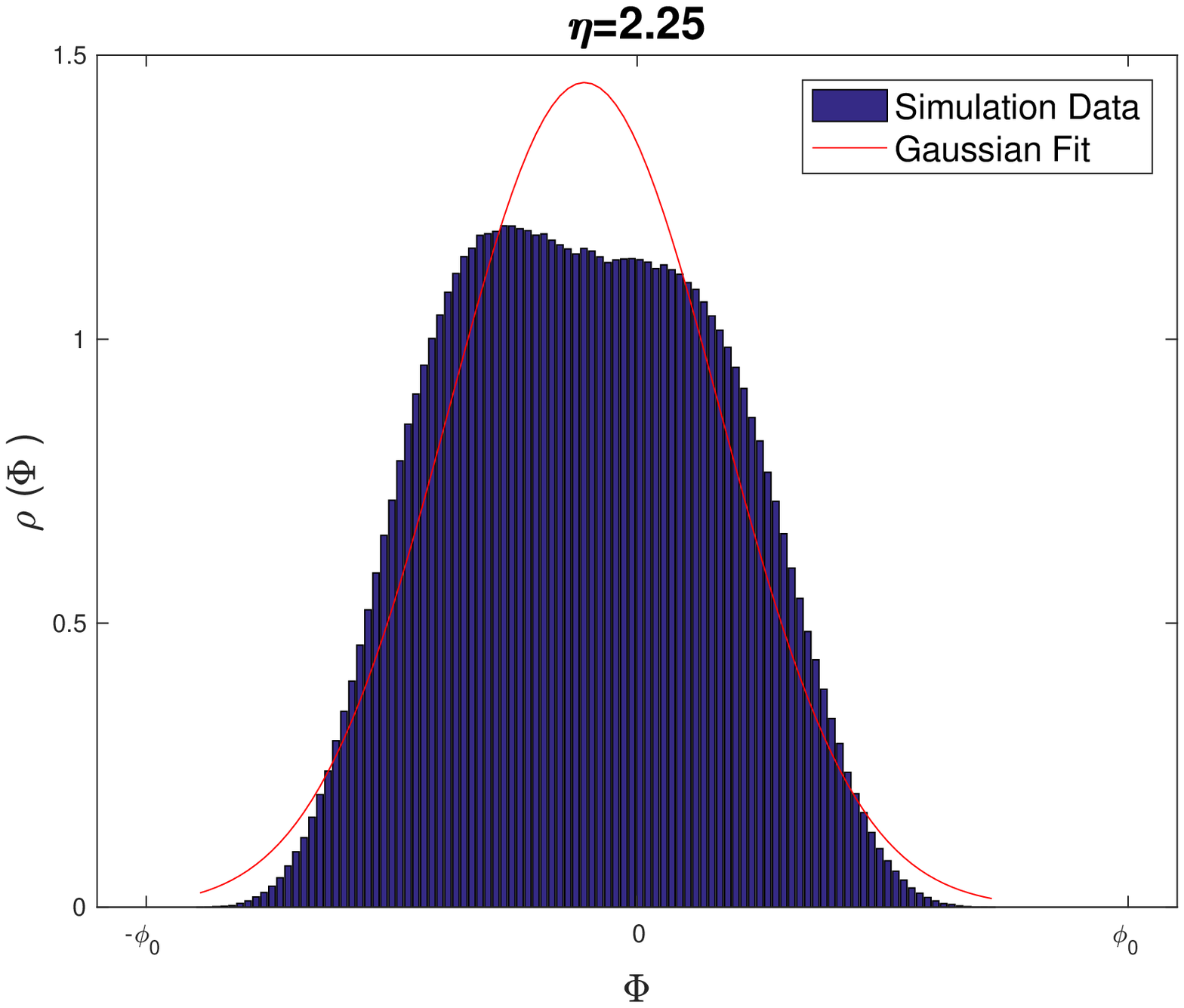}
\includegraphics[width=5cm]{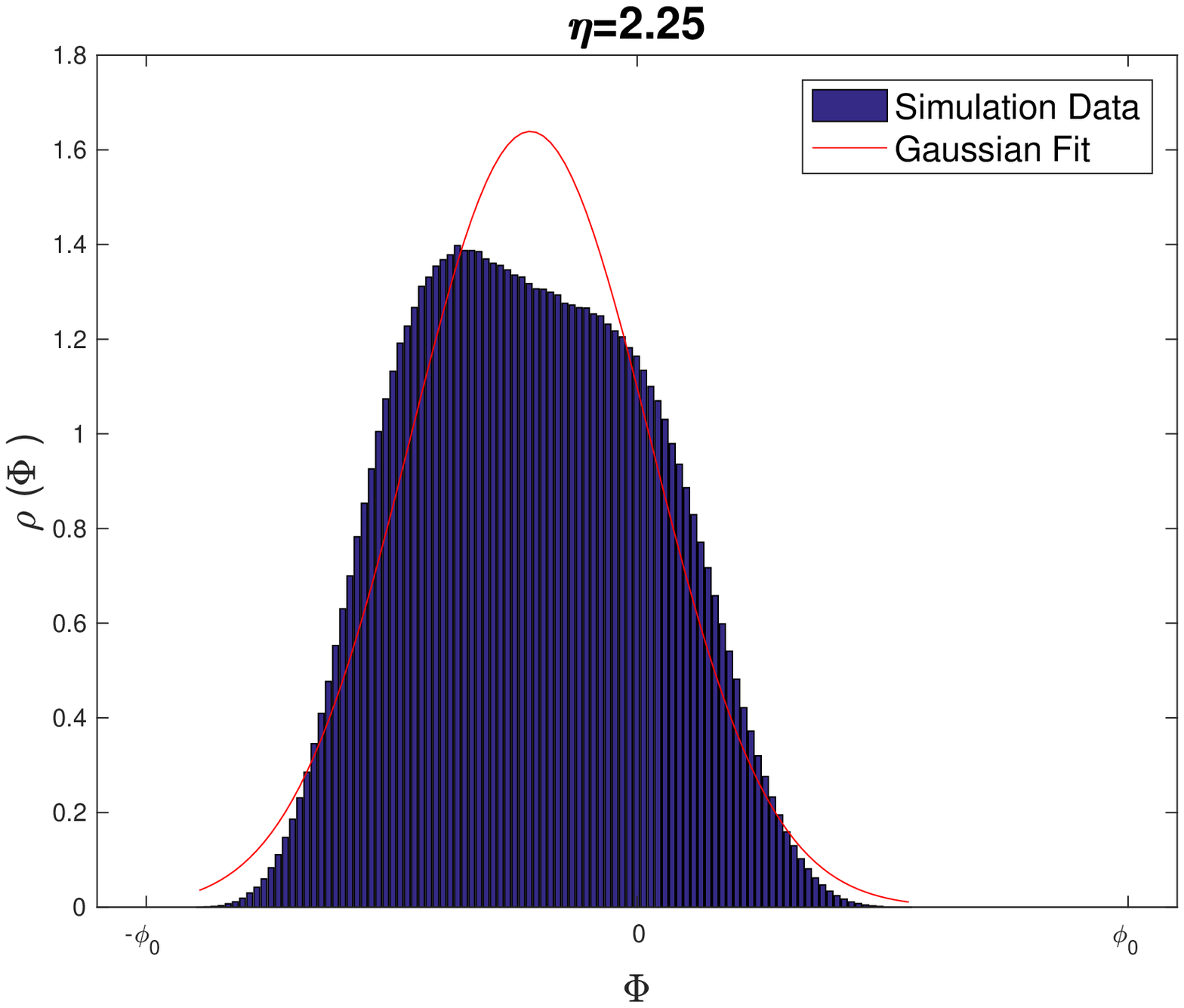}
\includegraphics[width=5cm]{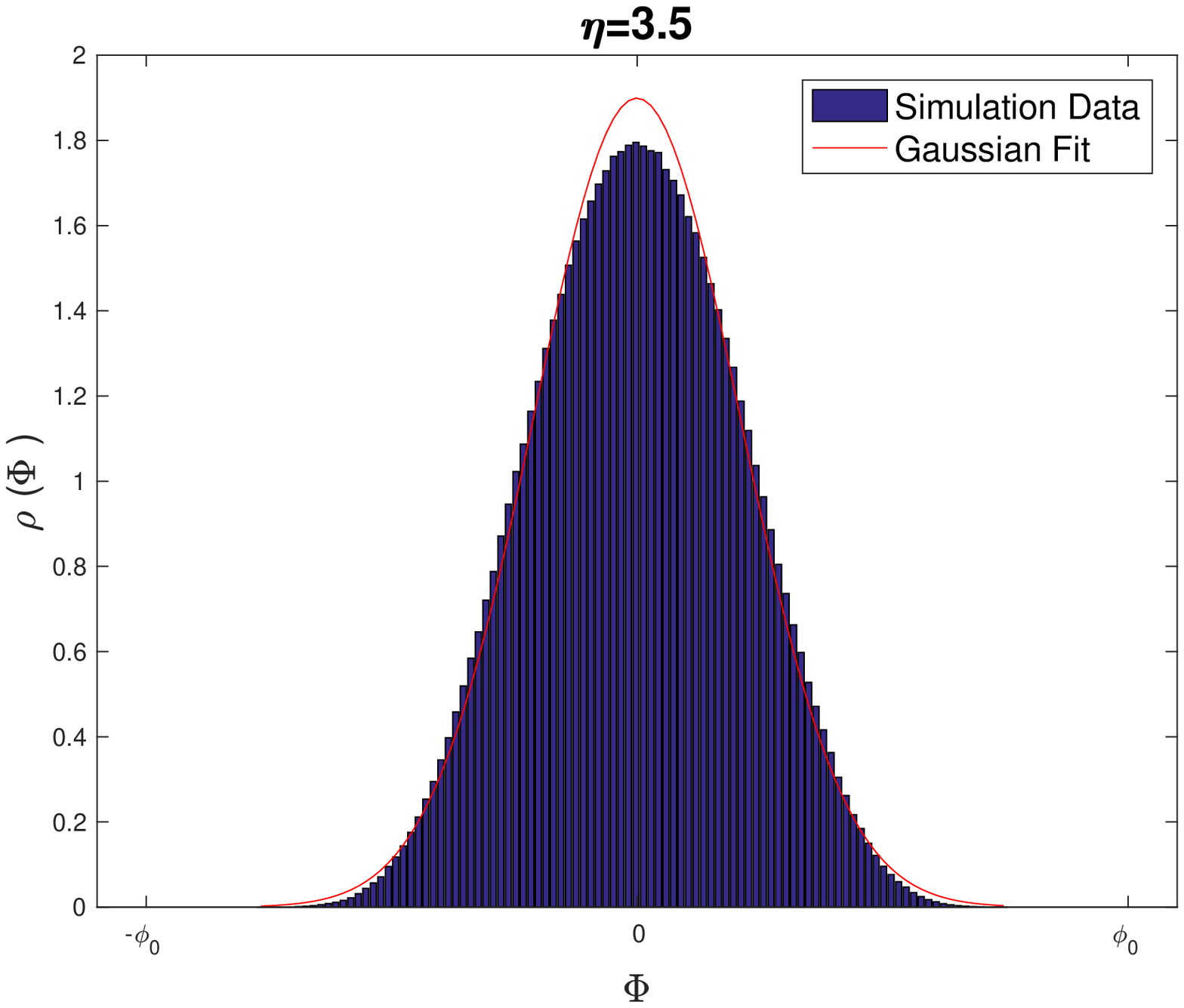}
\includegraphics[width=5cm]{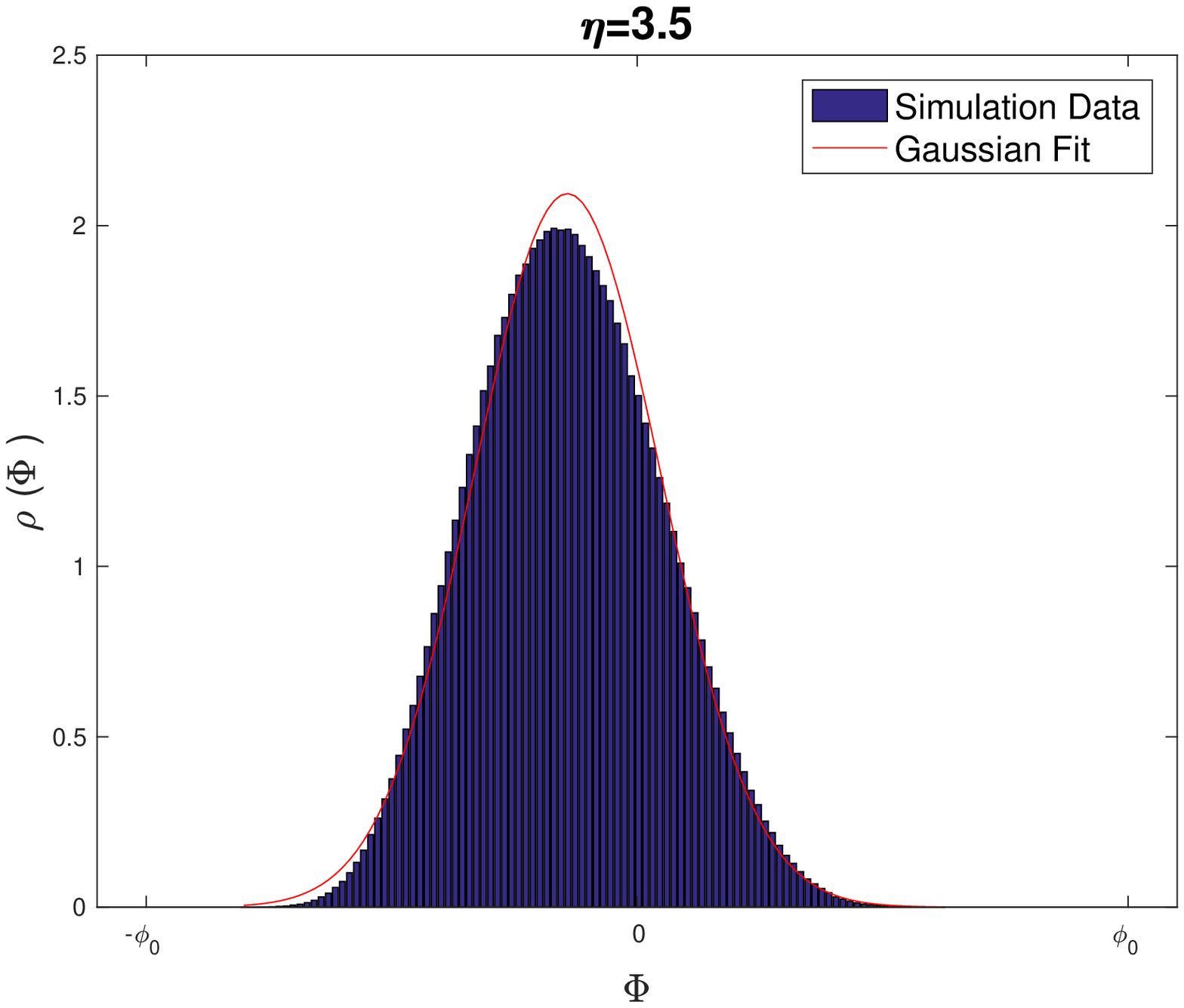}
\includegraphics[width=5cm]{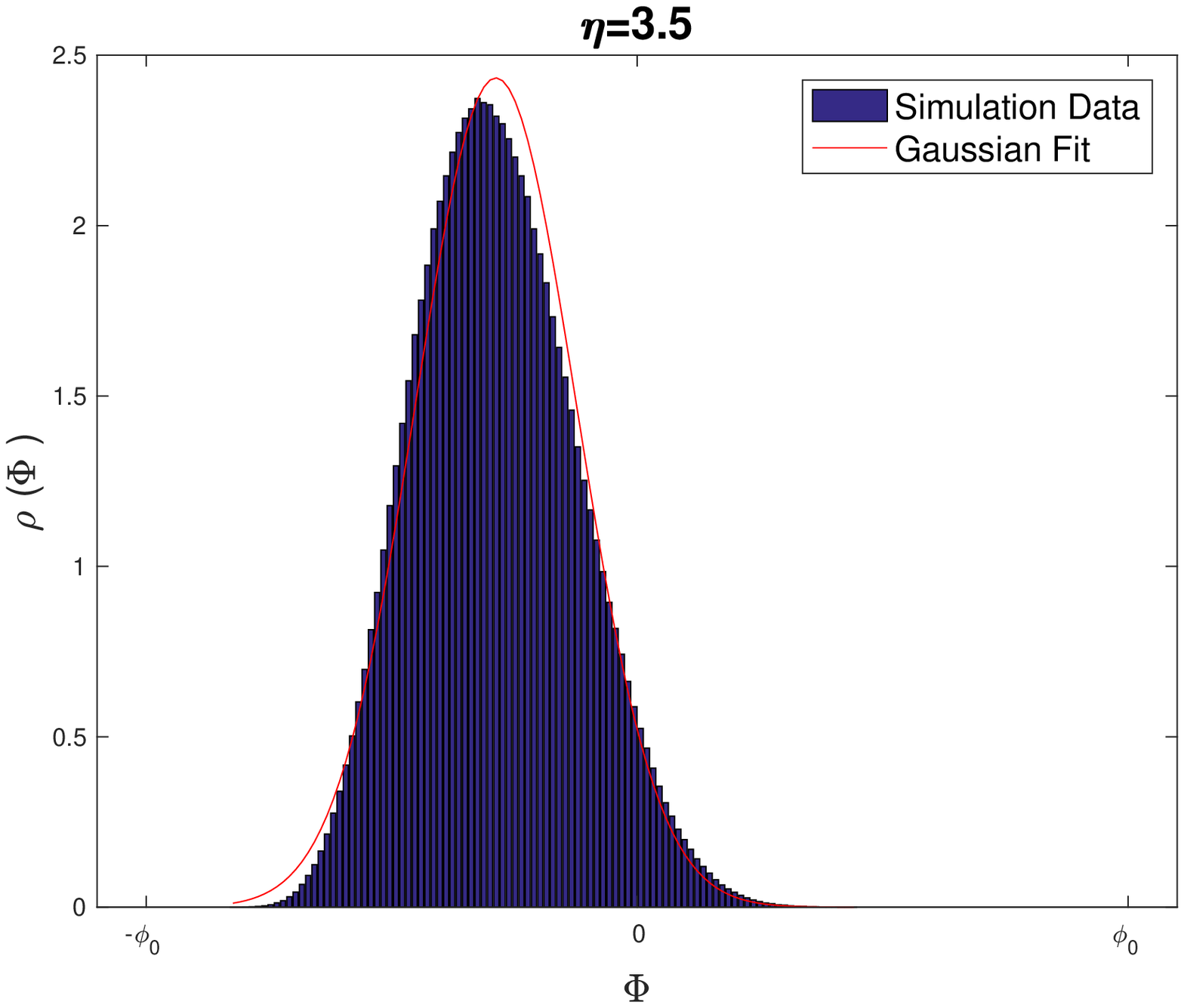}
\includegraphics[width=5cm]{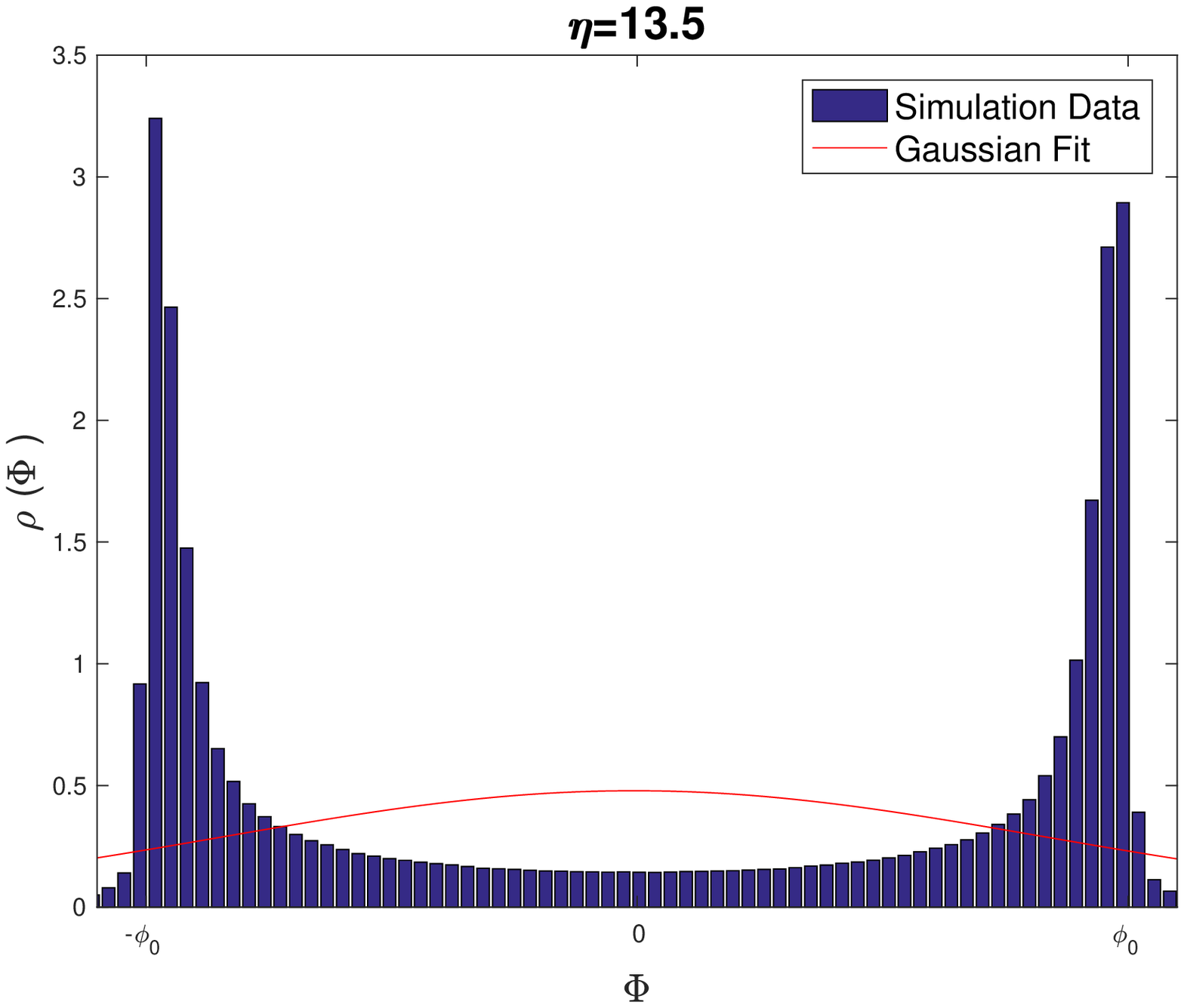}
\includegraphics[width=5cm]{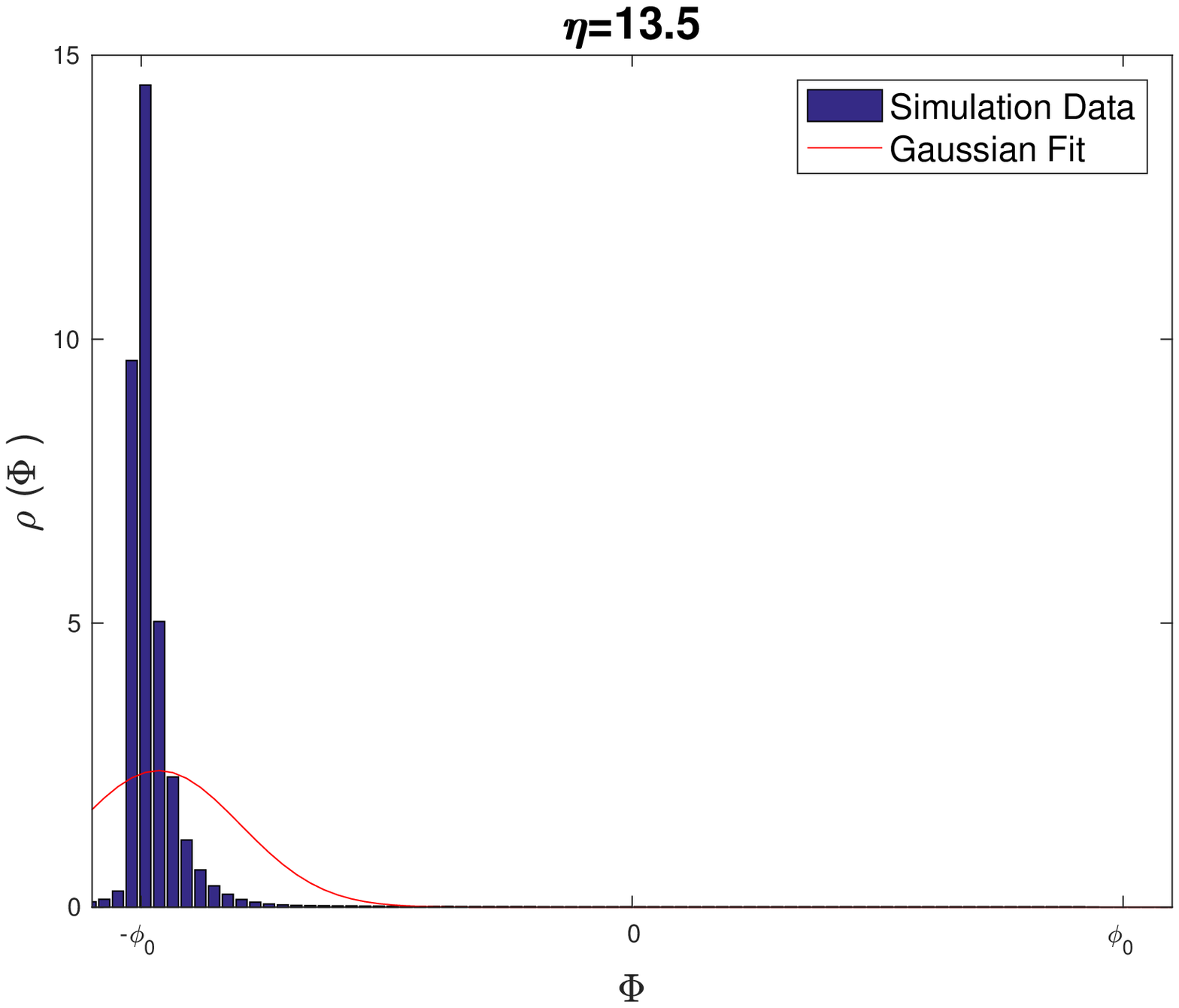}
\includegraphics[width=5cm]{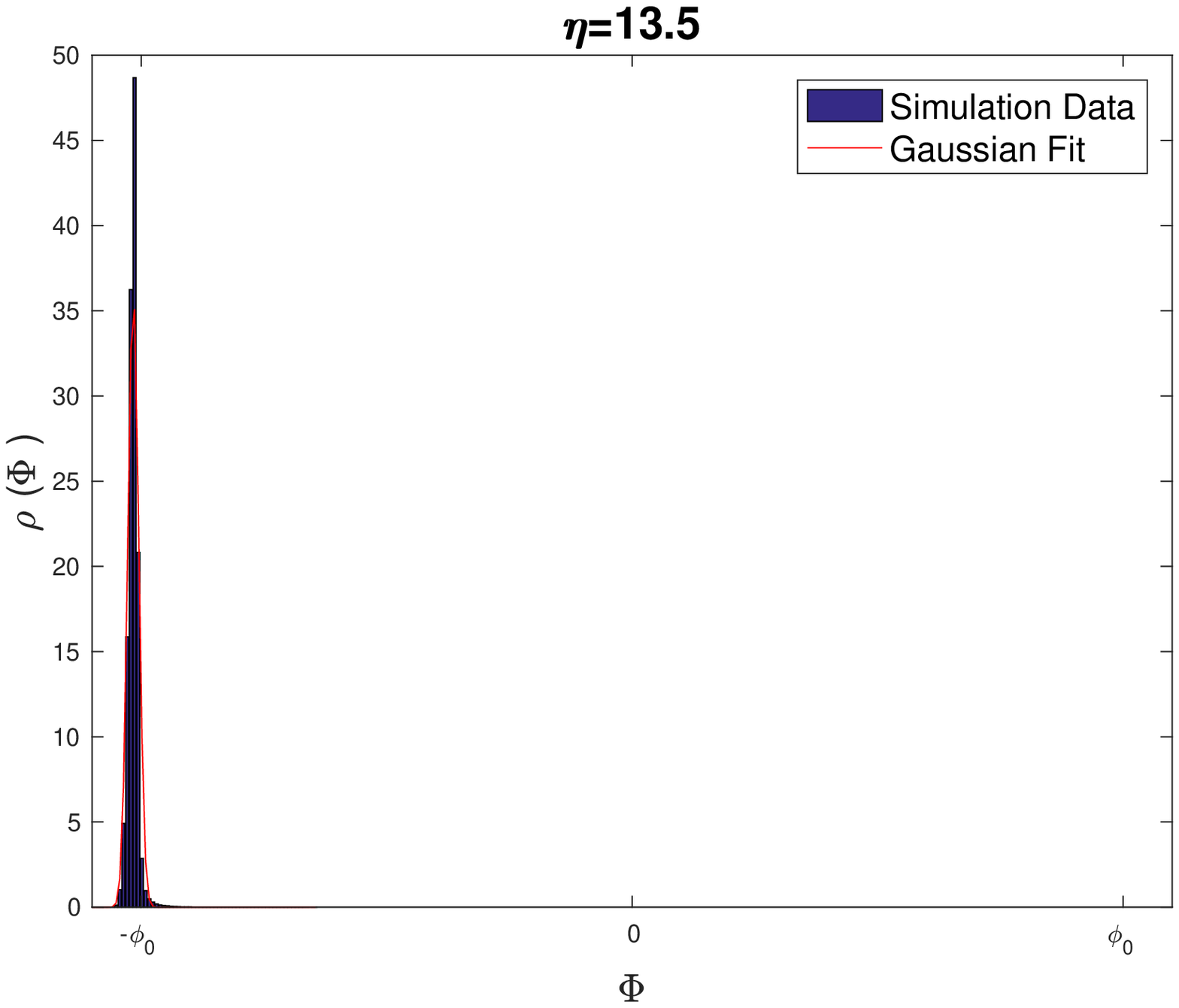}
\caption{\label{figure1}Distribution of the field $\phi$ at the conformal times $\eta=2.25$ (top row panels), $\eta=3.5$ (middle row panels) and $\eta=13.5$ (bottom row panels), for different initial population biases ($b=1$ in the left column panels, $b=0.8$ in the middle column panels, and $b=0.6$ in the right column panels).}
\end{figure*}
\begin{figure*}
\includegraphics[width=7cm]{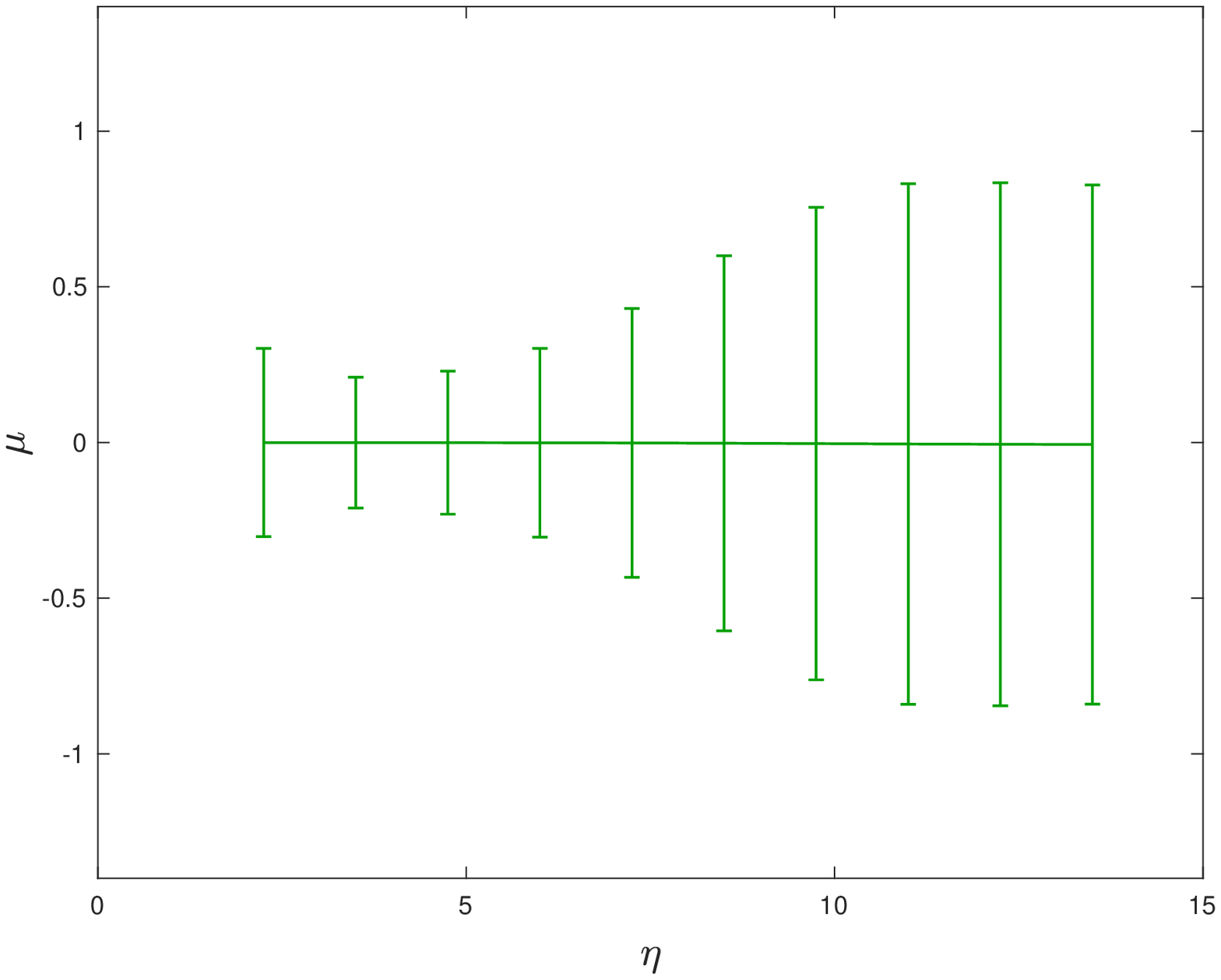}
\includegraphics[width=7cm]{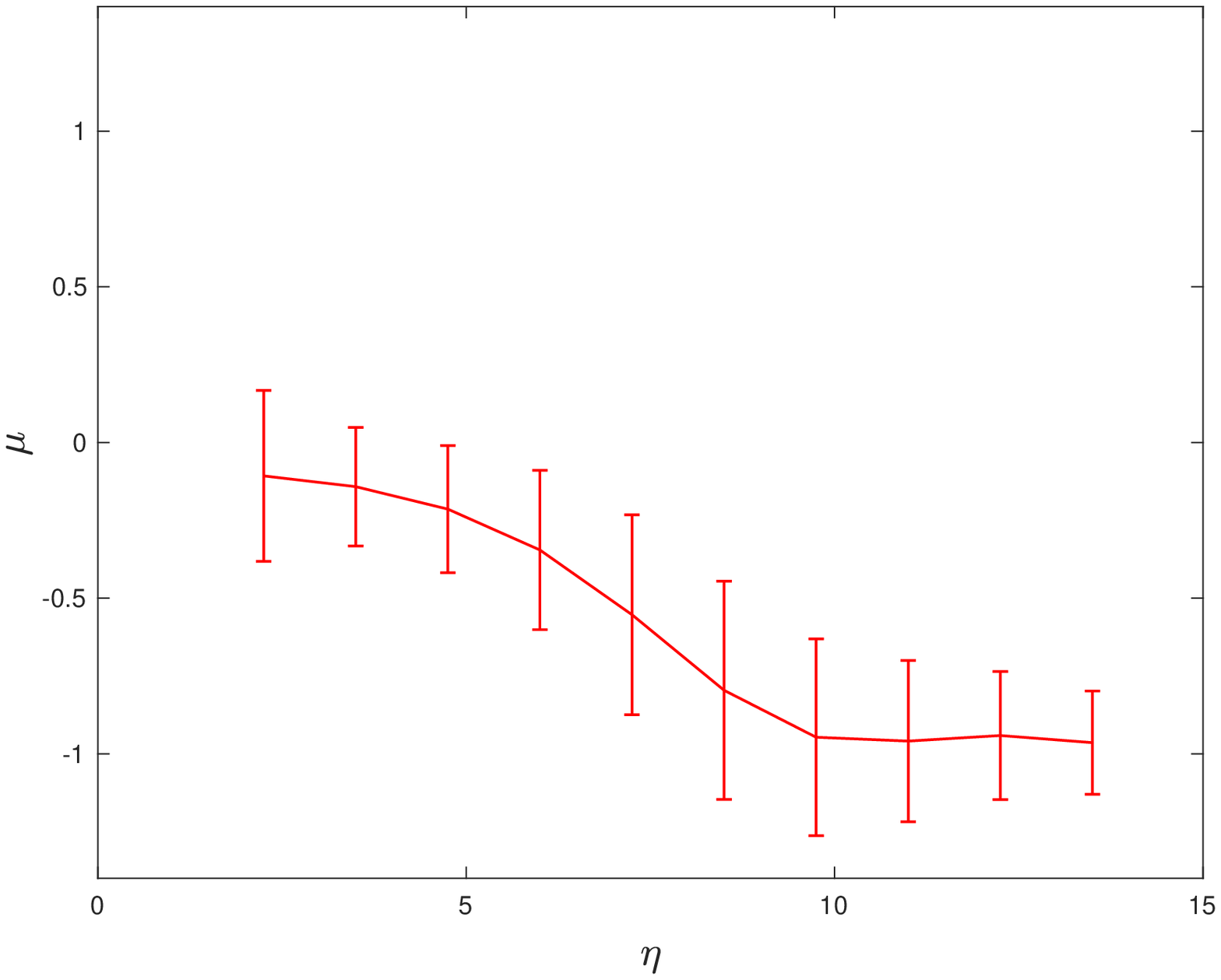}
\vskip0in
\includegraphics[width=7cm]{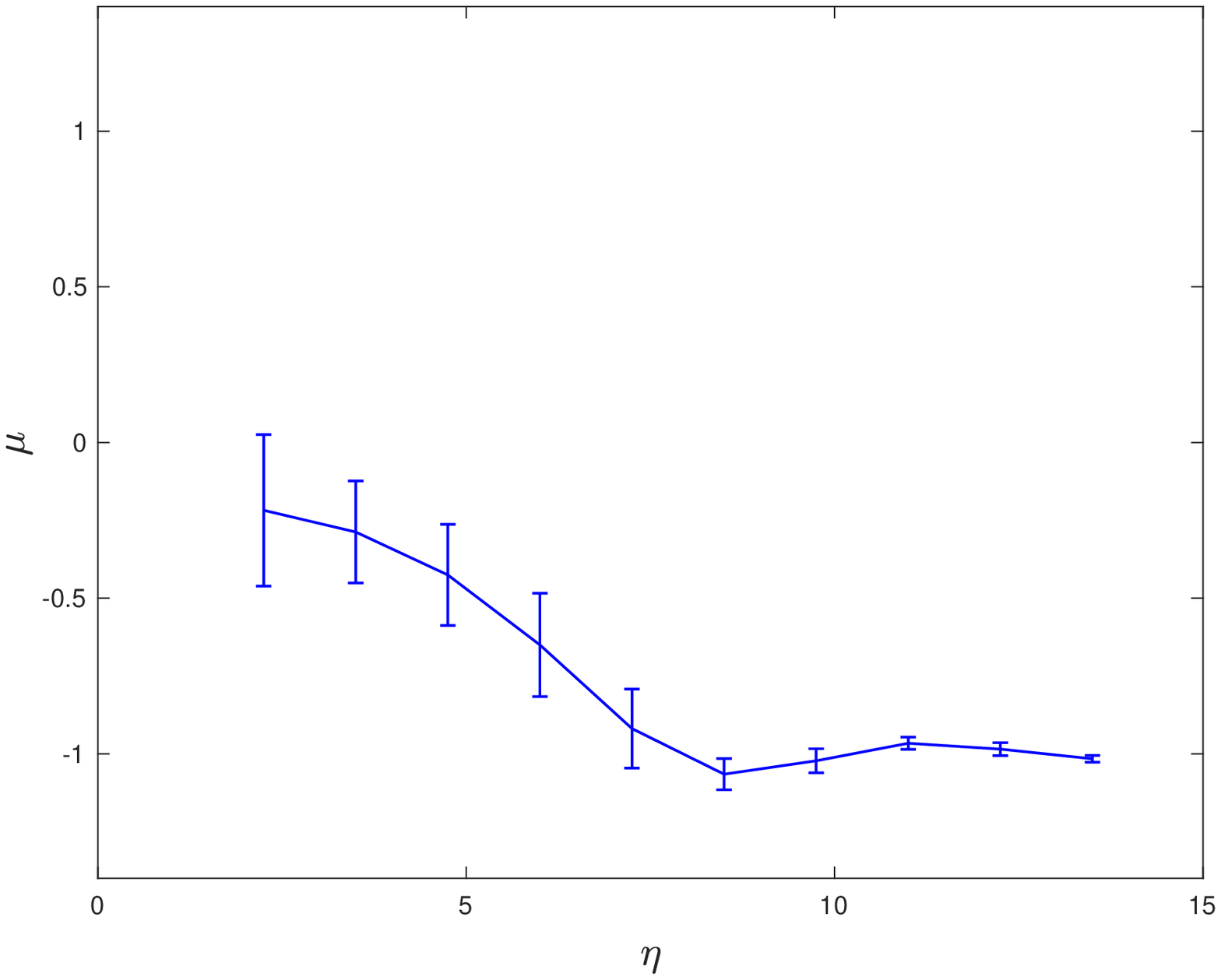}
\includegraphics[width=7cm]{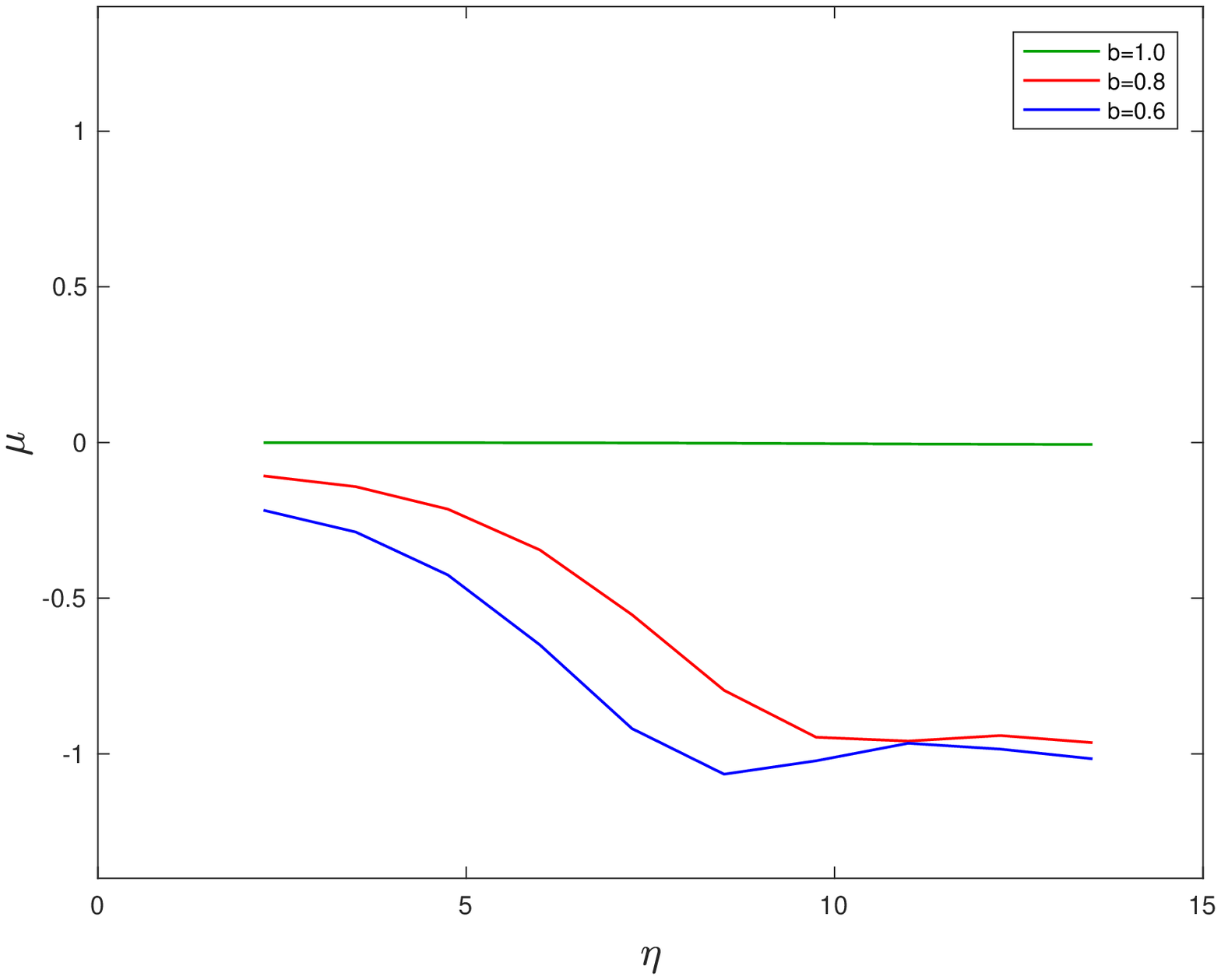}
\caption{\label{figure2}The evolution of the mean, $\mu$, of the Gaussian fit obtained from the field distribution at various timesteps in the biased initial conditions case. The error bars depict the standard deviation. The simulation parameters are respectively $b=1$ (top left panel), $b=0.8$ (top right) and $b=0.6$ (bottom left). The bottom right panel directly compares the evolution of the mean values in these three cases.}
\end{figure*}


\section{\label{biases}Biases and Gaussianity}

In order to test our Gaussianity hypothesis, we carried out a statistical analysis of the number of cells of each given simulation box (in this case, boxes of size $2048^2$), by producing histograms of the frequency of each value of the field, and performing a Gaussian fit to this distribution. We repeated this procedure for various timesteps, and studied how the distributions change as the simulations evolve and the networks disappear.

For the case of biased initial conditions we have done this analysis for three different initial population biases: $b=1.0$, $b=0.8$ and $b=0.6$; the first is unbiased, while the others, respectively, represent the cases of weak and strong bias. Figure \ref{figure1} depicts this diagnostic for three different timesteps, showing that the effects of the bias are distinguishable at very early stage ($\eta=2.25$, top row panels), especially if we look at the $b=0.6$ case. The evolution of this bias can be observed in Fig. \ref{figure2} as well, where the mean $\mu$ of the fit to the field distribution is skewed at a very early stage for the biased cases. 

We note that in the unbiased case the Gaussian approximation is reasonable around time $\eta=3.5$ (cf. the middle-row left-hand-side panel). The larger the bias, the worse the agreement between the Gaussian fit and the field distribution, and the more the mean is shifted from $\phi=0$. This leads us to conclude that the Gaussian assumption is not warranted for these decays. It is worthy of note that the Gaussian regime in the unbiased case occurs at the timesteps with minimal standard deviation, as expected; this standard deviation subsequently increases, as a larger fraction of the box is in the minima of the potential. On the other hand, in the biased cases (and especially with a strong bias) the standard deviation decreases as the network decays. Looking at the last plotted timestep (corresponding to $\eta=13.5$, bottom row of Fig. \ref{figure1}) we also confirm that in the unbiased case all points end up equally divided between the two minima, whereas in the biased cases all points end up in one of the minima. This behavior is also clear by looking at the evolution of the mean $\mu$ itself, in the last panel of Fig. \ref{figure2}.

\begin{figure*}
\includegraphics[width=5cm]{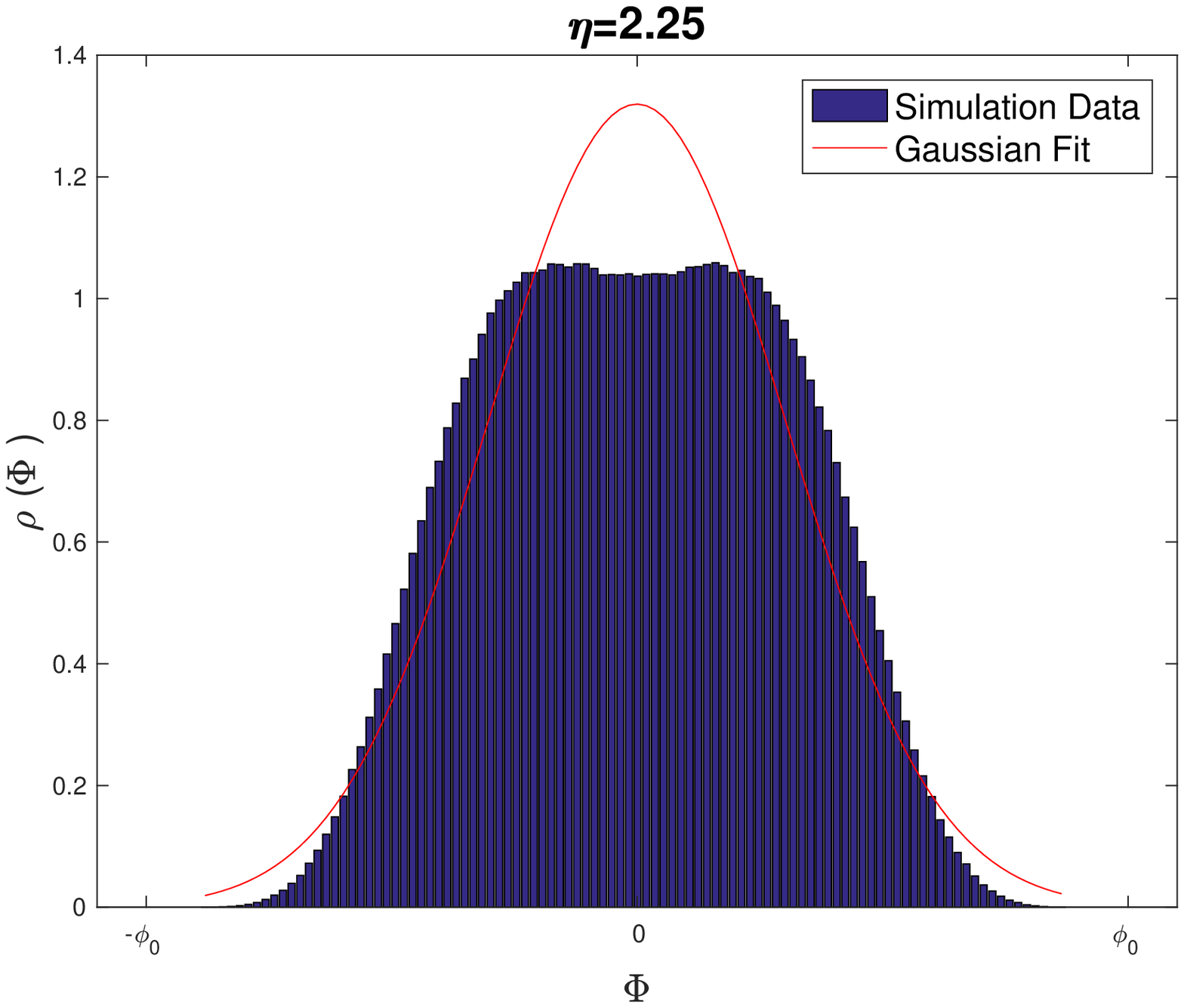}
\includegraphics[width=5cm]{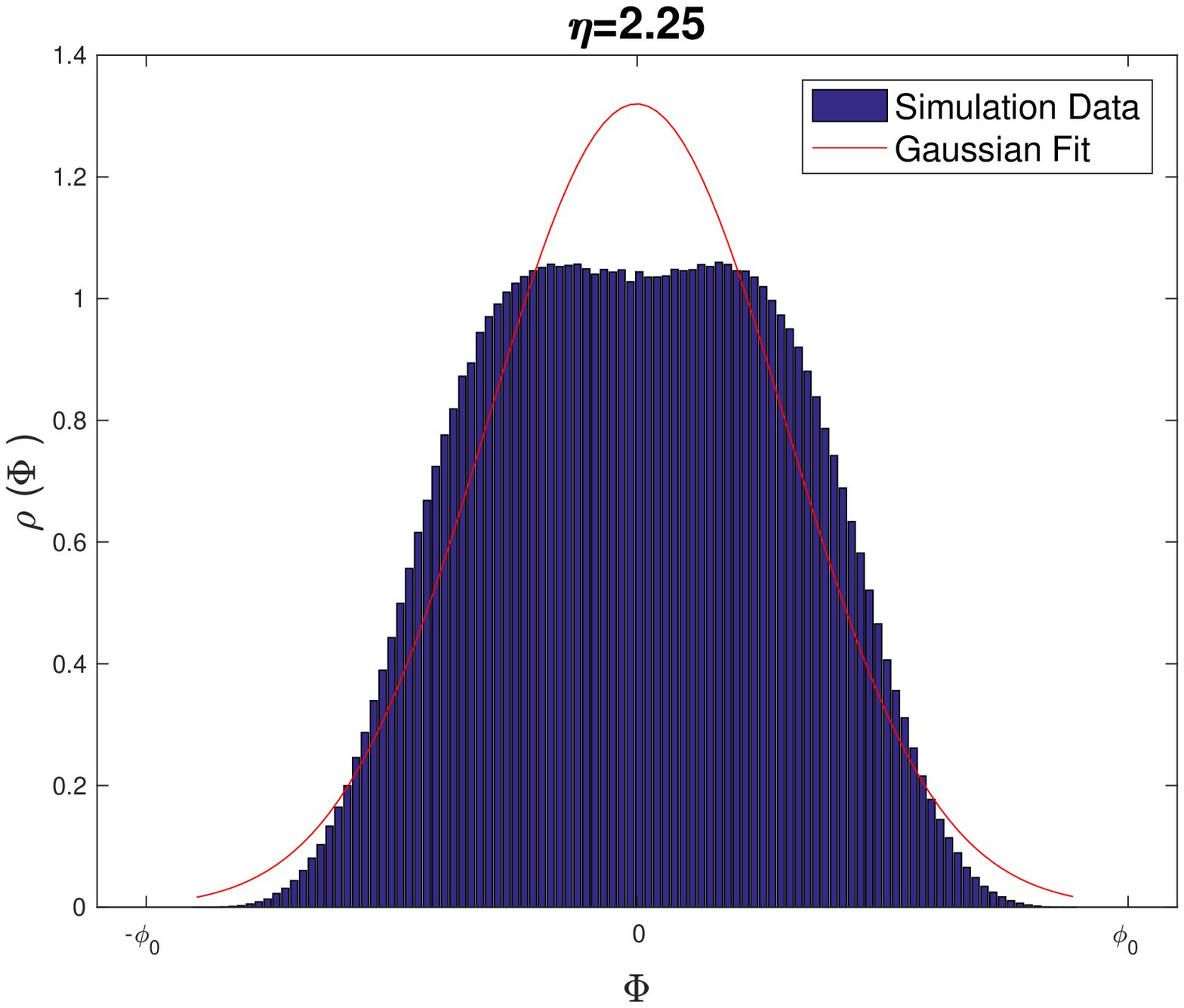}
\includegraphics[width=5cm]{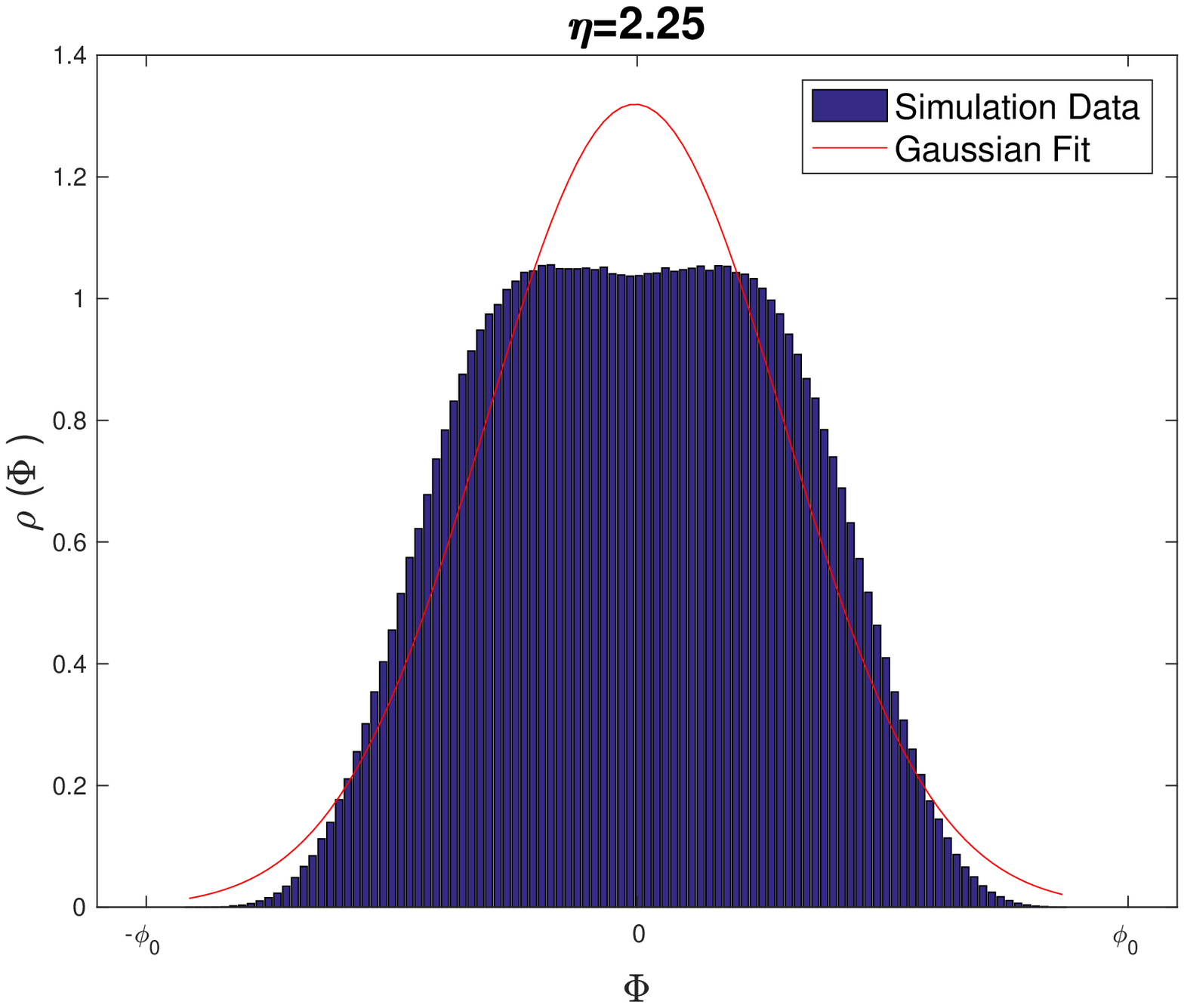}
\includegraphics[width=5cm]{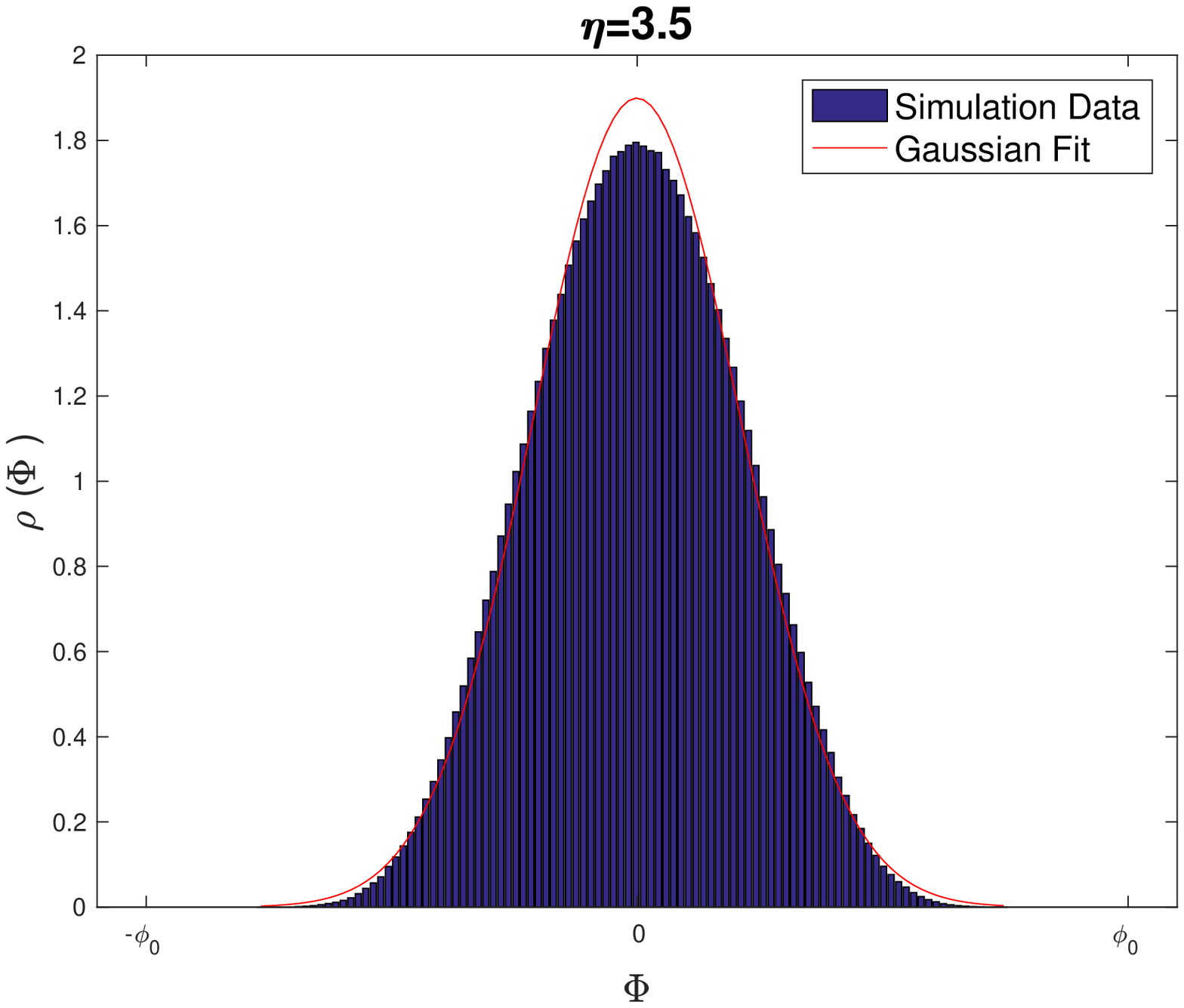}
\includegraphics[width=5cm]{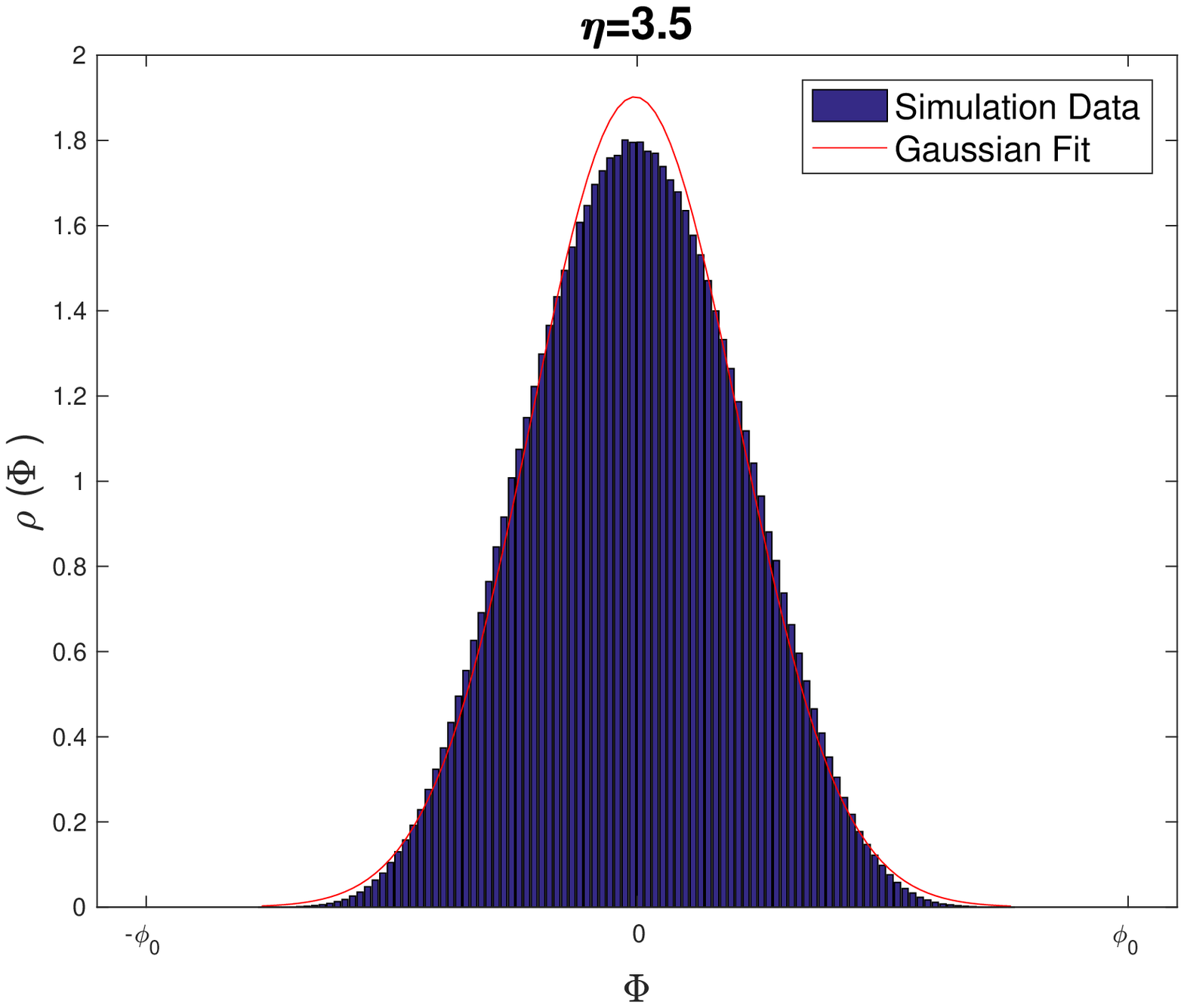}
\includegraphics[width=5cm]{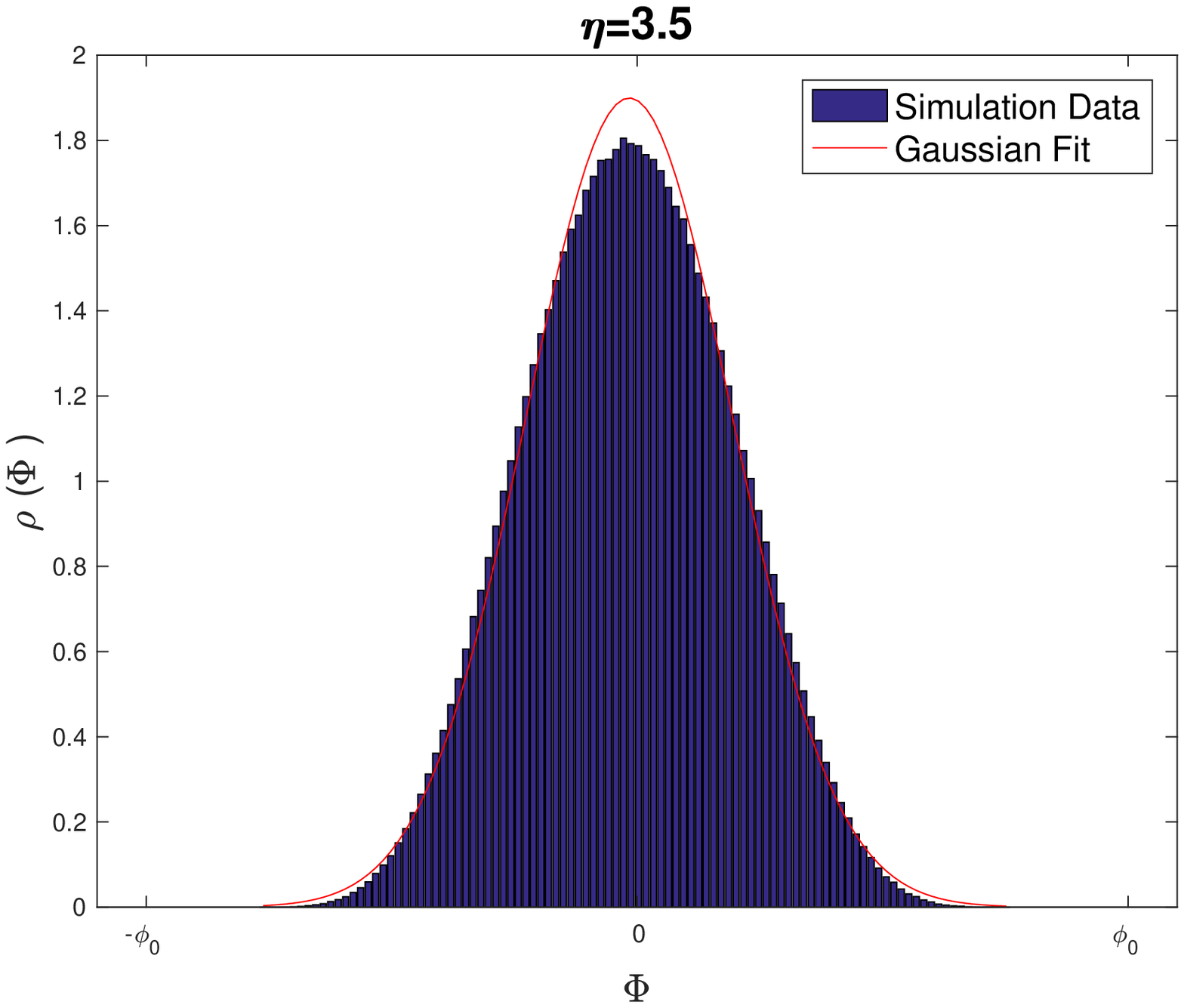}
\includegraphics[width=5cm]{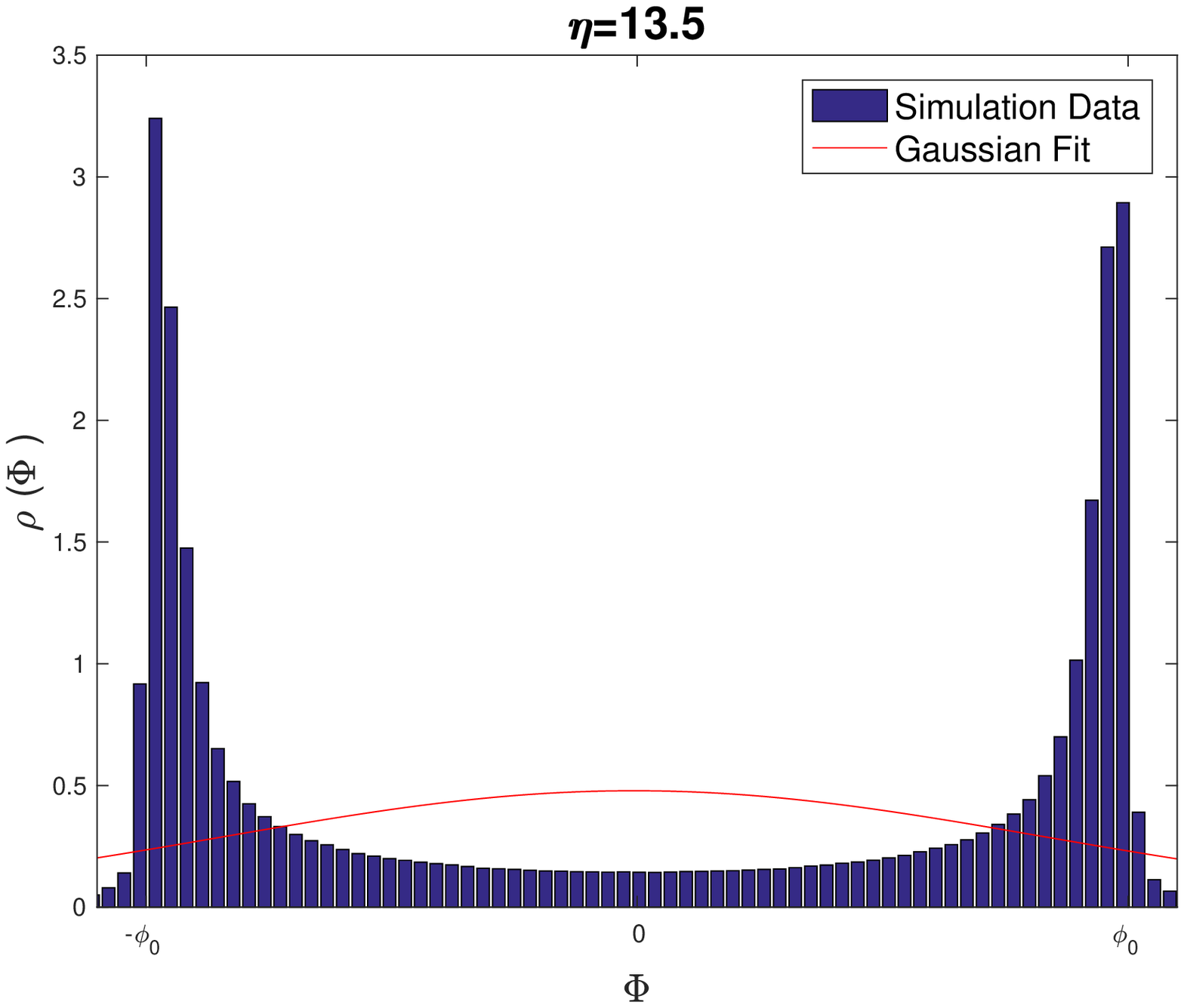}
\includegraphics[width=5cm]{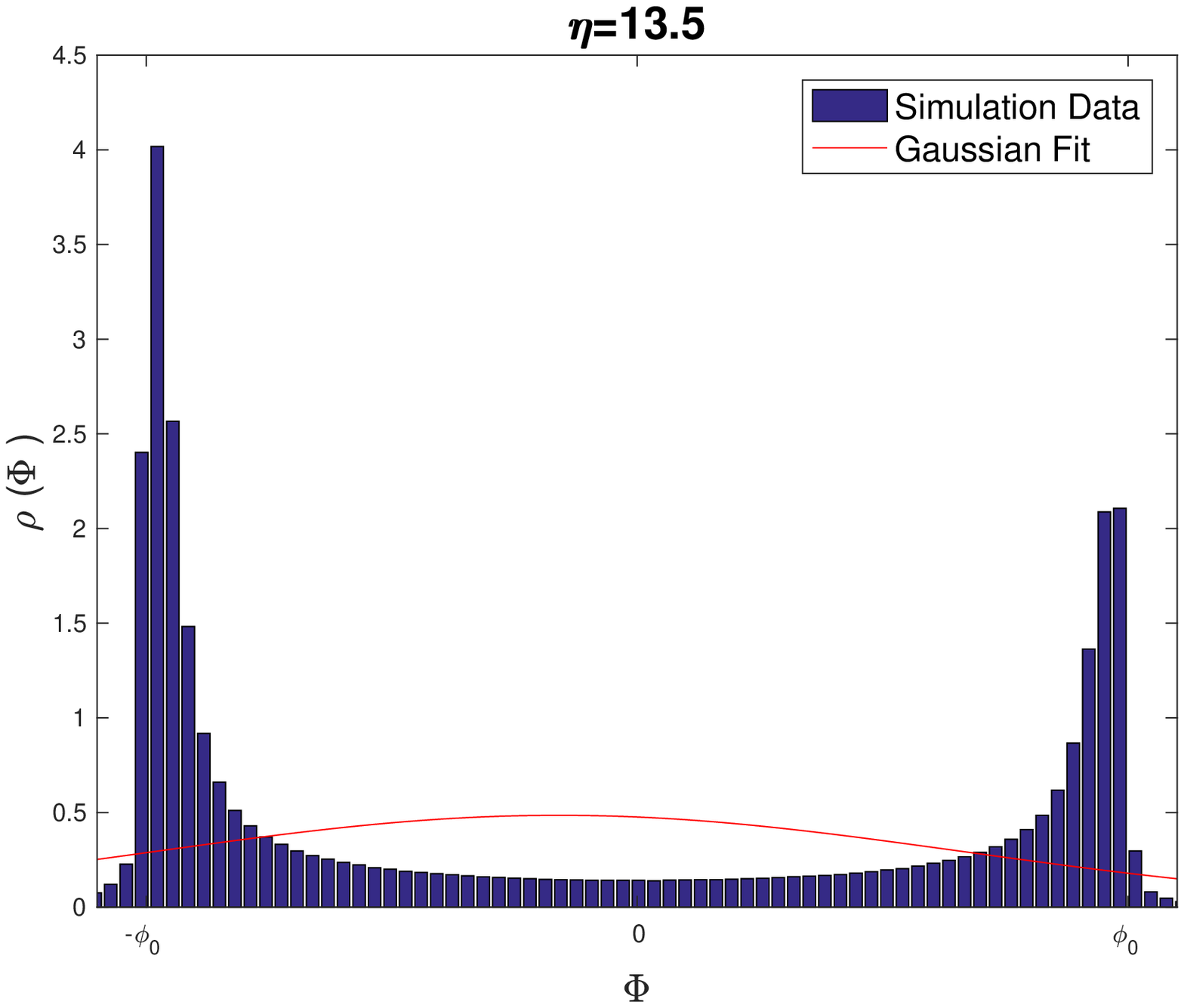}
\includegraphics[width=5cm]{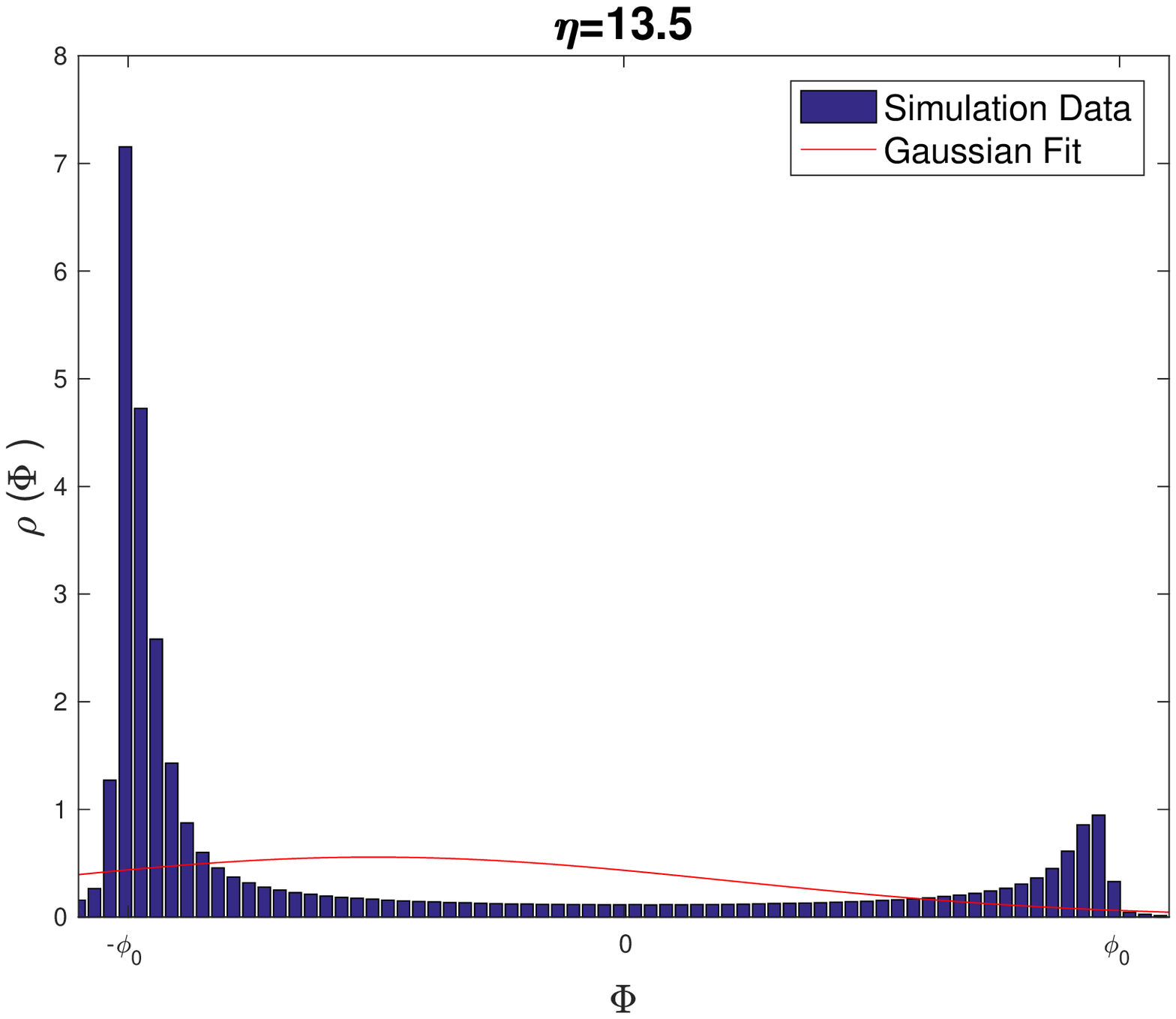}
\caption{\label{figure3}Distribution of the field $\phi$ at the conformal times $\eta=2.25$ (top row of panels), $\eta=3.5$ (middle row of panels) and $\eta=13.5$ (bottom row of panels), for different potential biases ($\theta = 0.0$ at the left column, $\theta=0.03$ at the middle column, and $\theta=0.1$ at the right column).}
\end{figure*}
\begin{figure*}
\includegraphics[width=7cm]{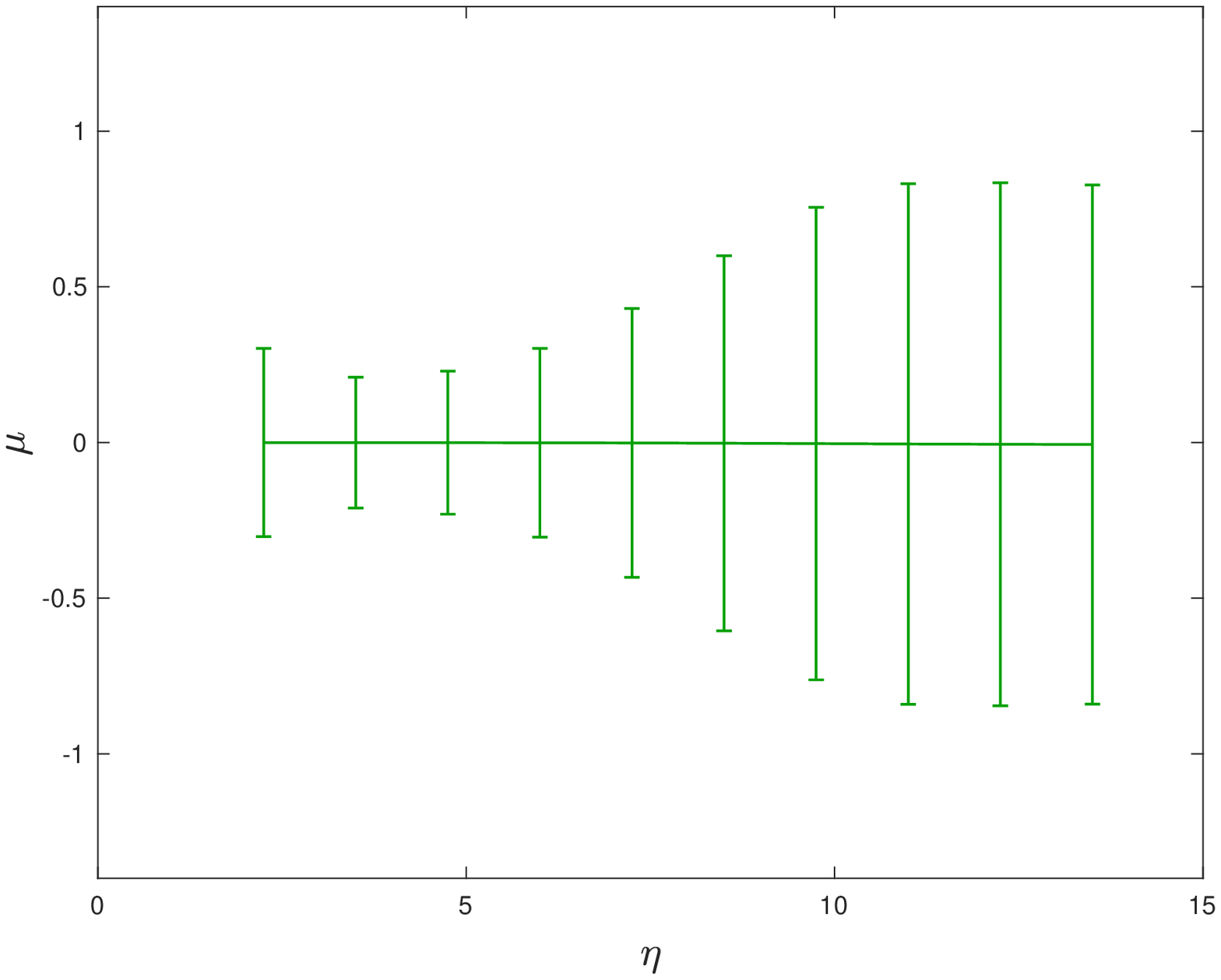}
\includegraphics[width=7cm]{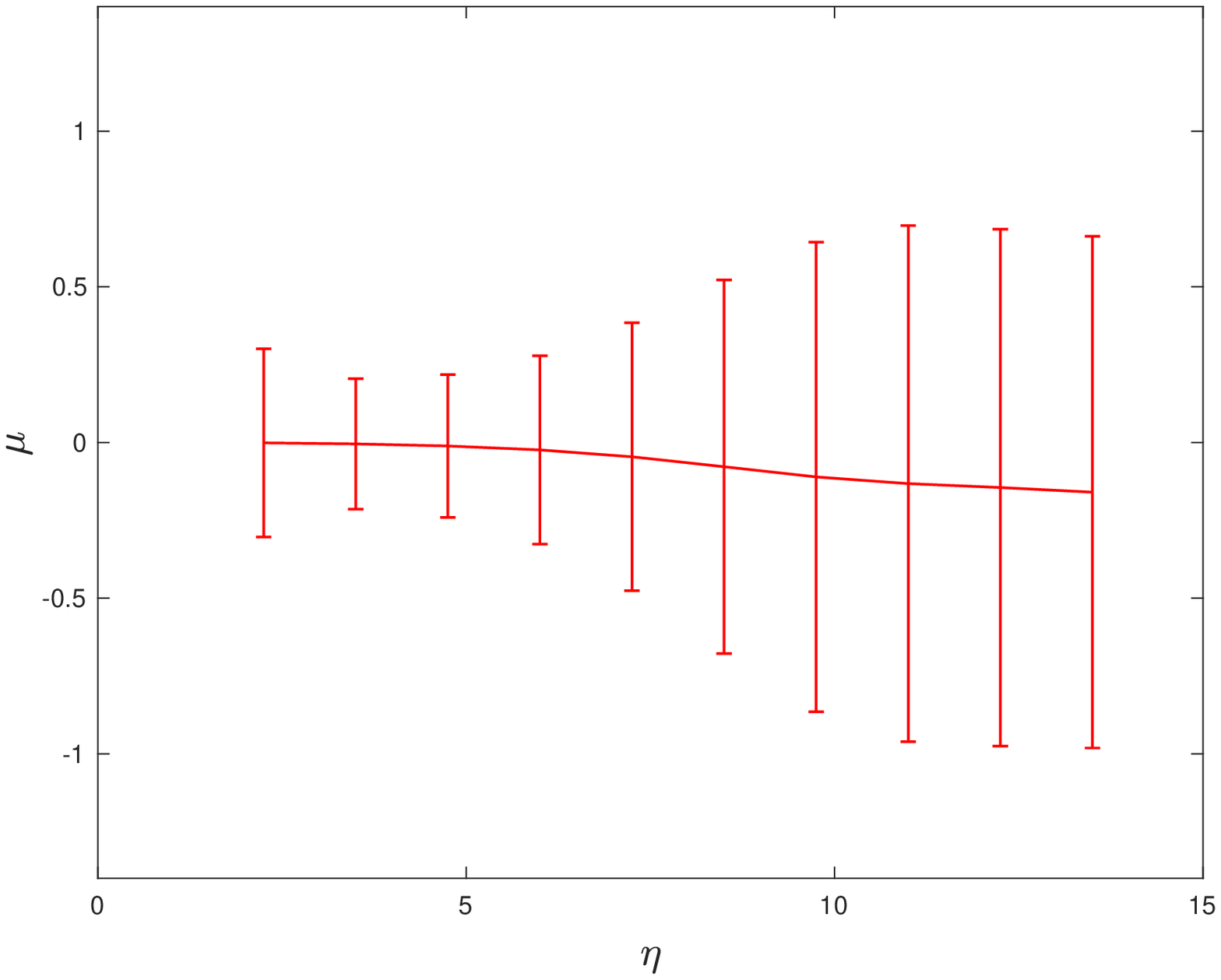}
\vskip0in
\includegraphics[width=7cm]{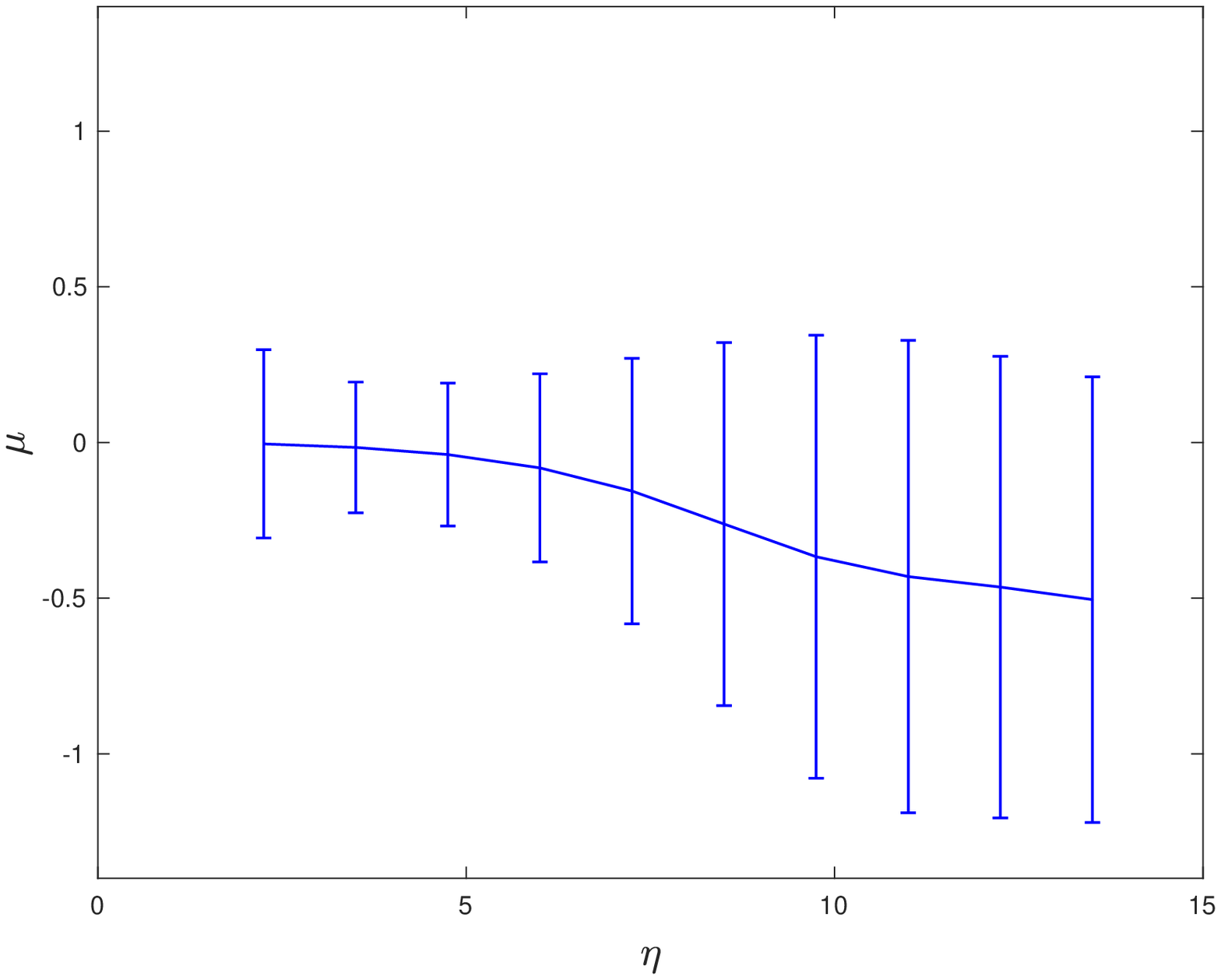}
\includegraphics[width=7cm]{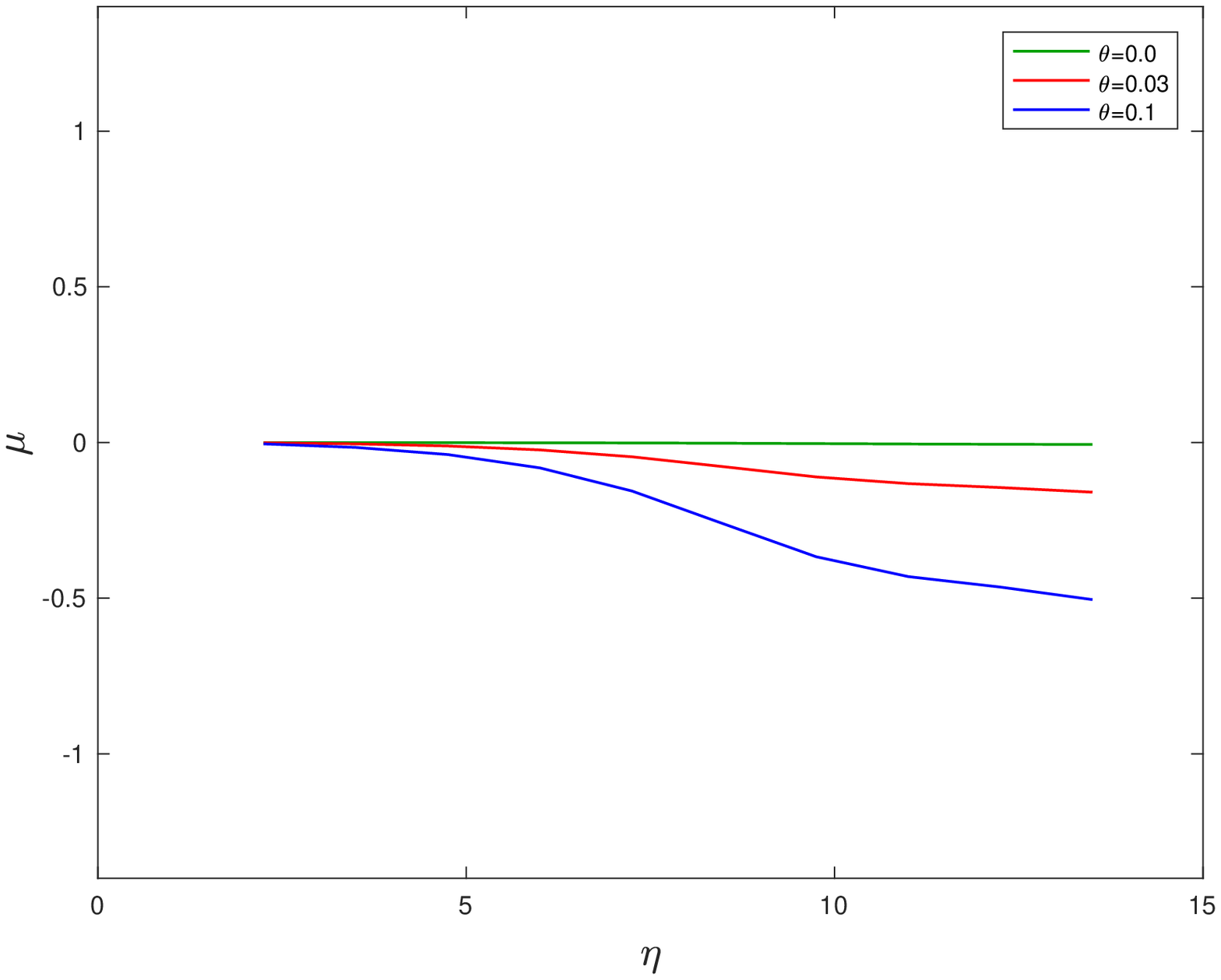}
\caption{\label{figure4}The evolution of the mean, $\mu$, of the Gaussian fit obtained from the field distribution at various timesteps in the biased potential case. The error bars depict the standard deviation. The simulation parameters are respectively $\theta=0$ (top left panel), $\theta=0.03$ (top right) and $\theta=0.1$ (bottom left). The bottom right panel directly compares the evolution of the mean values in these three cases.}
\end{figure*}

For the case of the biased potential the physical parameter being changed is the height difference between the potential's two minima, and we have done an analogous study for the three cases $\theta=0.0$ (for which there is no bias), $\theta = 0.03$ and $\theta= 0.1$.

Figure \ref{figure3} depicts this diagnostic for three different timesteps in this case. When compared to the previous case, the differences are clear. As can be seen at the $\eta=3.5$ timestep (cf. the middle row of panels), all biased potential networks now obey the Gaussian approximation at some range of (numerically) intermediate timesteps, when the network has erased its numerical initial conditions and is still evolving as in the standard scenario. As discussed in the previous section, this occurs because at early times the wall surface tension dominates, and as long as this is the case the network will have the standard behavior. Notice how, unlike the biased initial conditions case, our walls exhibit little to no population bias at earlier times, $\eta=2.25$ (cf. the top row of panels). At later timesteps (for example $\eta=13.5$, cf. the bottom row of panels), the bias is quite prominent, as expected. This corresponds to the regime where the volume pressure is dominant. 

This is also clear by looking at the evolution of the average, $\mu$, of the Gaussian fit for this case, which is shown in Fig. \ref{figure4}. Unlike the previous case, and since the wall networks reasonably obey the Gaussian distribution approximation, the lowest standard deviations are expected at earlier timesteps---an expectation which is again confirmed. Thus our analysis fully supports our hypothesis, formulated in Paper I, that the Gaussianity approximation which is one of the assumptions leading to the Hindmarsh decay formula, is the key difference between the two decay mechanisms.


\section{\label{anisot}Anisotropy and isotropization}

As mentioned previously, the VOS model for domain walls was been qualitatively derived in \cite{VOSwalls}, and a more rigorous ab initio derivation was found by \cite{Rybak1}. (See also \cite{VOSbook} for a general discussion of the model.) It relies on two averaged quantities, a density $\rho_w$ (or equivalent a characteristic physical lengthscale $L$, related to the former via $\rho_w=\sigma/L$) and a root mean squared velocity $v$. In a FLRW, these evolve as
\begin{equation}
\frac{dL}{dt}=(1+3v^2)HL+c_wv
\end{equation}
\begin{equation}
\frac{dv}{dt} = (1-v^2) \bigg( \frac{k_w}{L} -3Hv \bigg)\,,
\end{equation}
where $c_w$ and $k_w$ are respectively the energy loss and momentum (or curvature) parameters. The model has been shown to be in very good agreement with high-resolution field theory simulations of domain wall networks in FLRW isotropic universes, for a vary broad range of expansion rates, but as larger and more accurate simulations were done (and different expansion rates were probed), it became clear that $c_w$ and $k_w$ are not constant but depend on velocity. This led to an extension of the original model \cite{Rybak1,Rybak2} Here, as another form of validation, we test whether the extended VOS model can predict the scaling regime of walls created in eras not necessarily isotropic.

A detailed derivation of the velocity dependence of the two model parameters can be found in \cite{Rybak1}. Here we will simply state the final result, which is relevant for our analysis. The momentum parameter has the form
\begin{equation}
    k(v) = k_0 \cdot\frac{1-(qv^{2})^{\beta_w}} {1+(qv^{2})^{\beta_w}}
\end{equation}
where $k_{0}$, $q$ and $\beta_w$ are constant parameters to be determined by comparison with simulations. The first of these represents the maximum value of the momentum parameter, and $k(v)$ approaches this value at small velocities, corresponding to high expansion rates. The second is the averaged maximal velocity for wall networks, therefore $0 < 1/q \leq v^2 \approx 2/3$, and $k(v)$ approaches this value in the opposite limit of low enough expansion rates. Finally $\beta_w$ (not to be confused with the expansion rate $\beta$) describes how fast the parameter interpolates between the two limiting cases. As for the energy loss term, $c_wv$ is complemented by an additional term
\begin{equation}
F(v)= c_{w}v + d[k_0 -k(v)]^{r}\,,
\end{equation}
where, $d$ and $r$ are two more constant parameters that model energy losses due to scalar radiation. With these extensions the VOS model equations have the following form, written in terms of the density $\rho_w$,
\begin{equation}
    \frac{d\rho_w}{d\eta}= -3v^2{\cal H}\rho_w - \frac{F(v)\rho^2_w}{\sigma}
\end{equation}
\begin{equation}
    \frac{dv}{d\eta} = (1-v^2) \left[\frac{k(v)\rho}{\sigma} -3{\cal H}v \right]\,;
\end{equation}
we have also expressed them in terms of conformal time $\eta$ (with ${\cal H}$ being the conformal Hubble parameter), for easier comparison with the numerical simulations. 

\begin{table*}
\begin{tabular}{c | c | c c |  c c }
    Case & $\beta $ & $\mu$ & $\nu$ & $\sigma/(\rho_w\tau)$ & $\gamma v$ \\
    \hline
A  & 1/2 & $-0.972 \pm 0.004$ & $-0.081 \pm 0.005$ & $0.547 \pm 0.018$ & $0.397 \pm 0.022$ \\
{} & 2/3 & $-0.973 \pm 0.013$ & $-0.043 \pm 0.008$ & $0.510 \pm 0.055$ & $0.338 \pm 0.021$ \\
{} & 4/5 & $-0.971 \pm 0.006$ & $-0.013 \pm 0.005$ & $0.410 \pm 0.020$ & $0.269 \pm 0.010$ \\
{} & 9/10 & $-1.024 \pm 0.006$ & $-0.028 \pm 0.006$ & $0.319 \pm 0.016$ & $0.192 \pm 0.009$ \\
{} & 95/100 & $-1.014 \pm 0.005$ & $0.022 \pm 0.006$ & $0.225 \pm 0.010$ & $0.136 \pm 0.006$ \\
{} & 99/100 & $-0.975 \pm 0.002$ & $0.010 \pm 0.003$ & $0.099 \pm 0.001$ & $0.059 \pm 0.001$ \\
    \hline
B  & 1/2 & $-0.985 \pm 0.010$ & $-0.017 \pm 0.006$ & $0.565 \pm 0.042$ & $0.408 \pm 0.017$ \\
{} & 2/3 & $-0.963 \pm 0.009$ & $-0.042 \pm 0.009$ & $0.495 \pm 0.038$ & $0.336 \pm 0.023$ \\
{} & 4/5 & $-1.031 \pm 0.006$ & $-0.049 \pm 0.005$ & $0.434 \pm 0.023$ & $0.271 \pm 0.012$ \\
{} & 9/10 & $-0.979 \pm 0.003$ & $-0.034 \pm 0.004$ & $0.305 \pm 0.007$ & $0.189 \pm 0.006$ \\
{} & 95/100 & $-0.992 \pm 0.003$ & $0.010 \pm 0.007$ & $0.220 \pm 0.006$ & $0.134 \pm 0.007$ \\
{} & 99/100 & $-0.990 \pm 0.002$ & $0.018 \pm 0.002$ & $0.100 \pm 0.001$ & $0.059 \pm 0.001$ \\
    \hline
C  & 1/2 & $-1.003 \pm 0.012$ & $-0.043 \pm 0.010$ & $0.504 \pm 0.046$ & $0.373 \pm 0.028$ \\
{} & 2/3 & $-0.959 \pm 0.008$ & $-0.037 \pm 0.006$ & $0.435 \pm 0.025$ & $0.316 \pm 0.014$ \\
{} & 4/5 & $-0.983 \pm 0.010$ & $-0.032 \pm 0.008$ & $0.376 \pm 0.028$ & $0.258 \pm 0.015$ \\
{} & 9/10 & $-0.987 \pm 0.008$ & $-0.046 \pm 0.006$ & $0.294 \pm 0.018$ & $0.187 \pm 0.009$ \\
{} & 95/100 & $-0.990 \pm 0.004$ & $0.029 \pm 0.005$ & $0.214 \pm 0.006$ & $0.135 \pm 0.006$ \\
{} & 99/100 & $-0.992 \pm 0.001$ & $0.021 \pm 0.003$ & $0.100 \pm 0.001$ & $0.059 \pm 0.001$ \\
    \hline
D  & 1/2 & $-0.992 \pm 0.010$ & $-0.012 \pm 0.007$ & $0.351 \pm 0.028$ & $0.307 \pm 0.017$ \\
{} & 2/3 & $-0.933 \pm 0.016$ & $-0.097 \pm 0.011$ & $0.320 \pm 0.039$ & $0.258 \pm 0.023$ \\
{} & 4/5 & $-0.962 \pm 0.005$ & $-0.043 \pm 0.009$ & $0.294 \pm 0.012$ & $0.230 \pm 0.016$ \\
{} & 9/10 & $-0.990 \pm 0.008$ & $-0.004 \pm 0.008$ & $0.250 \pm 0.014$ & $0.184 \pm 0.011$ \\
{} & 95/100 & $-0.971 \pm 0.005$ & $0.036 \pm 0.006$ & $0.191 \pm 0.007$ & $0.134 \pm 0.006$ \\
{} & 99/100 & $-0.982 \pm 0.002$ & $0.023 \pm 0.003$ & $0.097 \pm 0.002$ & $0.059 \pm 0.002$ \\
\end{tabular}
\caption{\label{tab:asym}Scaling exponents $\mu$ and $\nu$, and asymptotic values $\sigma/(\rho \tau)$ and  $\gamma v$ for each case mentioned in the text (A, B, C and D) for each of the simulated expansion rates. A fit range 501.25 - 3096.25 was used in all cases. One-sigma statistical uncertainties are shown throughout.}
\end{table*}

We begin by calculating asymptotic values for the two diagnostic quantities, for six different expansion rates in the range $0.5 < \beta < 0.99$. The results of this analysis are shown in Table \ref{tab:asym}.  A fit range of $\eta=501.25$ to $\eta=3096.25$ was used in all cases. At earlier times than the former the network is far from the scaling regimes, while at later times than the latter there are too few walls left in the simulation box to ensure good statistics. Nevertheless note that, as already discussed in Paper I, the more anisotropic a network is the longer it takes to reach the scaling regime. In particular, it is clear that the timescale for the evolution of the Case D simulations is slightly different from that of the other cases.

\begin{table*}[]
\begin{tabular}{c | c | c c c | c c c}
Case & $\beta$ & $c_w$ & $d$ & $r$ & $\beta_w$ & $k_0$ & $q$ \\
\hline
Case A & $0.5\le\lambda\le0.99$ & $0.00 \pm 0.01$  & $0.28 \pm 0.01$ & $1.44 \pm 0.12$ & $1.92 \pm 0.21$ & $1.71 \pm 0.01$ & $5.07 \pm 0.39$ \\
Case B & $0.5\le\lambda\le0.99$ & $-0.00 \pm 0.01$ & $0.29 \pm 0.02$ & $1.57 \pm 0.20$ & $1.47 \pm 0.18$ & $1.73 \pm 0.02$ & $4.08 \pm 0.43$ \\
Case C & $0.5\le\lambda\le0.99$ & $0.00 \pm 0.01$ & $0.23 \pm 0.03$ & $1.84 \pm 0.21$ & $1.37 \pm 0.17$ & $1.73 \pm 0.03$ & $5.25 \pm 0.52$ \\
Case D & $0.5\le\lambda\le0.99$ & $-0.00 \pm 0.01$ & $0.13 \pm 0.01$ & $2.05 \pm 0.22$ & $1.30 \pm 0.18$ & $1.71 \pm 0.05$ & $8.97 \pm 0.78$ \\
Global & $0.5\le\lambda\le0.99$ & $0.00 \pm 0.01$ & $0.22 \pm 0.02$ & $2.10 \pm 0.39$ & $1.52 \pm 0.30$ & $1.72 \pm 0.03$ & $5.68 \pm 0.89$ \\
\hline
Ref. \protect\cite{Rybak1}& $0.5\le\lambda\le0.9$ & $0.00 \pm 0.03$ & $0.29 \pm 0.01$ & $1.30 \pm 0.06$ & $1.65 \pm 0.17$ & $1.72 \pm 0.03$ & $4.10 \pm 0.17$ \\
Ref. \protect\cite{Rybak2}& $0.2\le\lambda\le0.9998$ & $0.00 \pm 0.08$ & $0.26 \pm 0.02$ & $1.42 \pm 0.04$ & $1.08 \pm 0.07$ & $1.77 \pm 0.03$ & $3.35 \pm 0.32$ \\
\end{tabular}
    \caption{\label{tab:param}Best fit parameters, with one-sigma statistical uncertainties, for the extended VOS for the $8192^2$ domain wall simulations in the present work (cases A, B, C and D, plus a global fit to all the data). For comparison, the bottom two rows show the parameters obtained for $4096^3$ simulations of standard walls in \protect\cite{Rybak1,Rybak2}.}
\end{table*}

We then use the same minimization procedure as in \cite{Rybak1,Rybak2} to obtain the best-fit VOS model parameters. The resulting parameters, together with the corresponding one-sigma uncertainties, are shown in Table \ref{tab:param}. We do this separately for each of the four cases being considered, and confirm that the best-fit parameters for Case D are slightly different from those of the other three cases. Nevertheless, we note the good agreement between all cases, bearing in mind that the reported error bars are one-sigma statistical uncertainties. (For a discussion of additional systematic uncertainties in these simulations, see \cite{Rybak2}.) Additionally we also show the results of a single global fit to all for sets of simulations. For comparison we also show the values obtained in the analysis of $4096^3$ simulations, reported in \cite{Rybak1,Rybak2}. While the speed of the approach to scaling depends both on the expansion rate $\beta$ and the stretch factor, and is clearly slower in the anisotropic cases, it is again clear that the best fit parameters are consistent throughout, given the aforementioned uncertainties. We note that in no case does the formation of wall blobs becomes a more efficient energy loss mechanism than scalar radiation, in other words $c_w=0$ is the best-fit parameter for all cases. We also note the remarkable consistency in the best-fit values of the parameter $k_0$.

\begin{figure*}
\includegraphics[width=8.5cm]{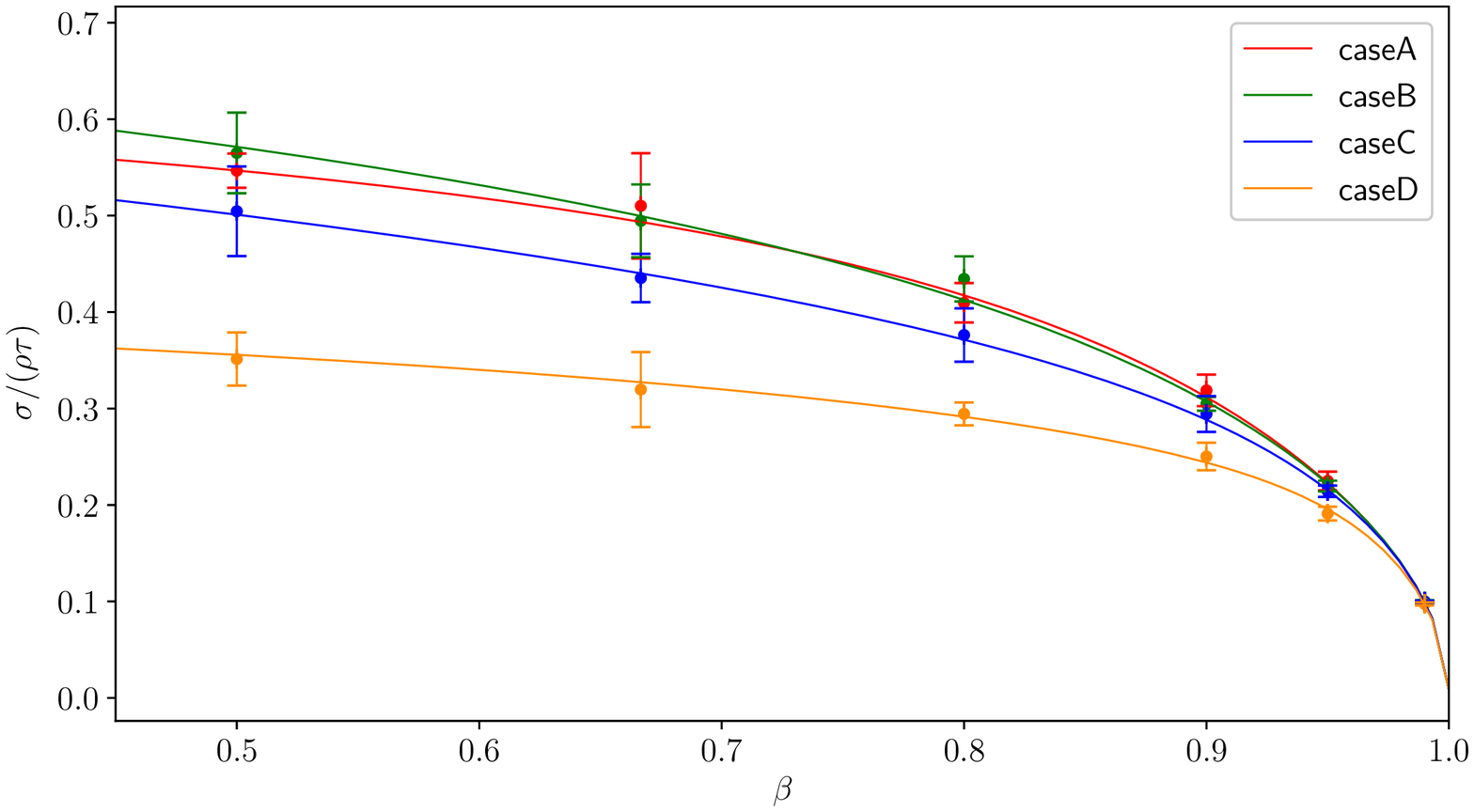}
\includegraphics[width=8.5cm]{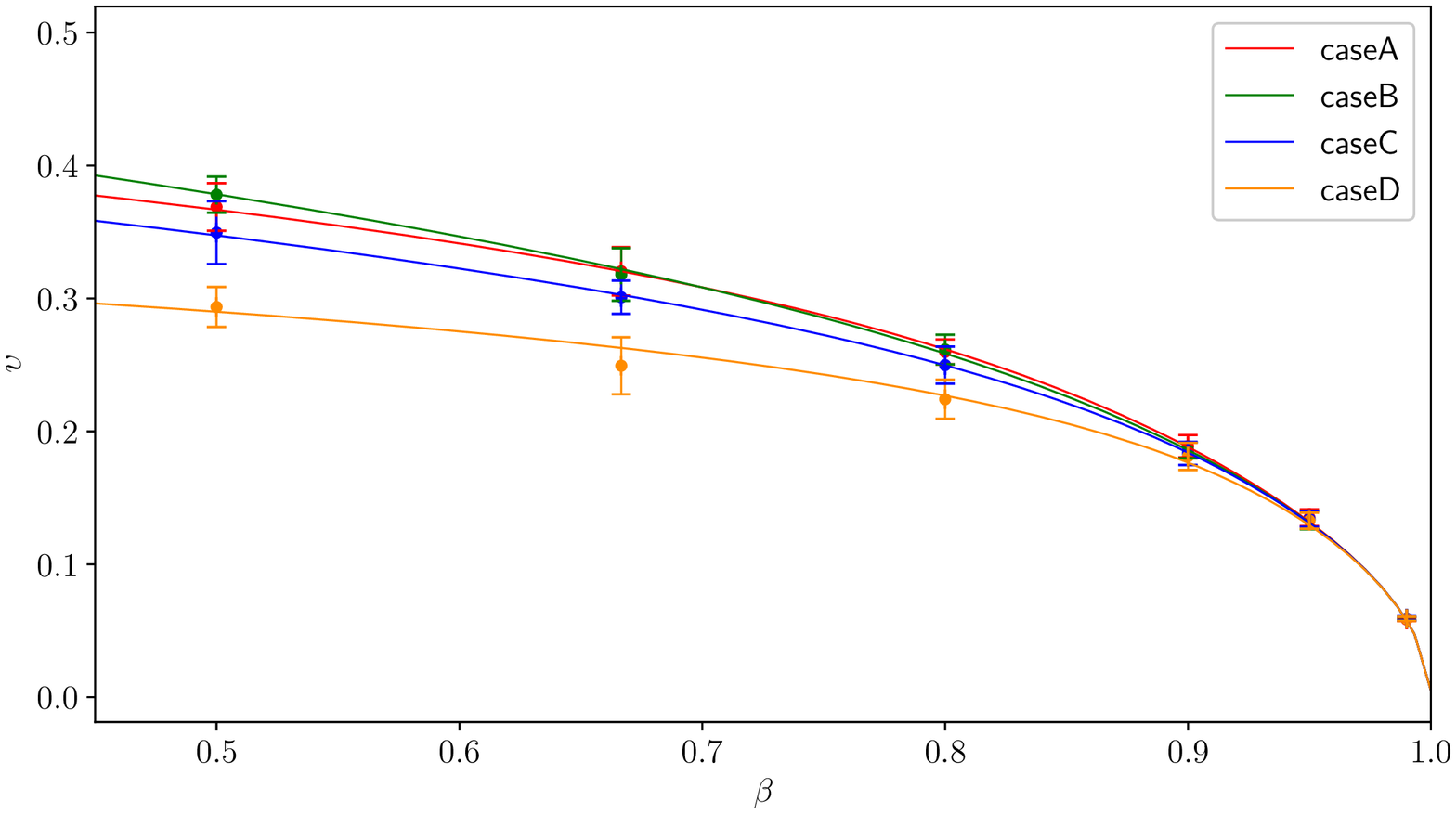}
\includegraphics[width=8.5cm]{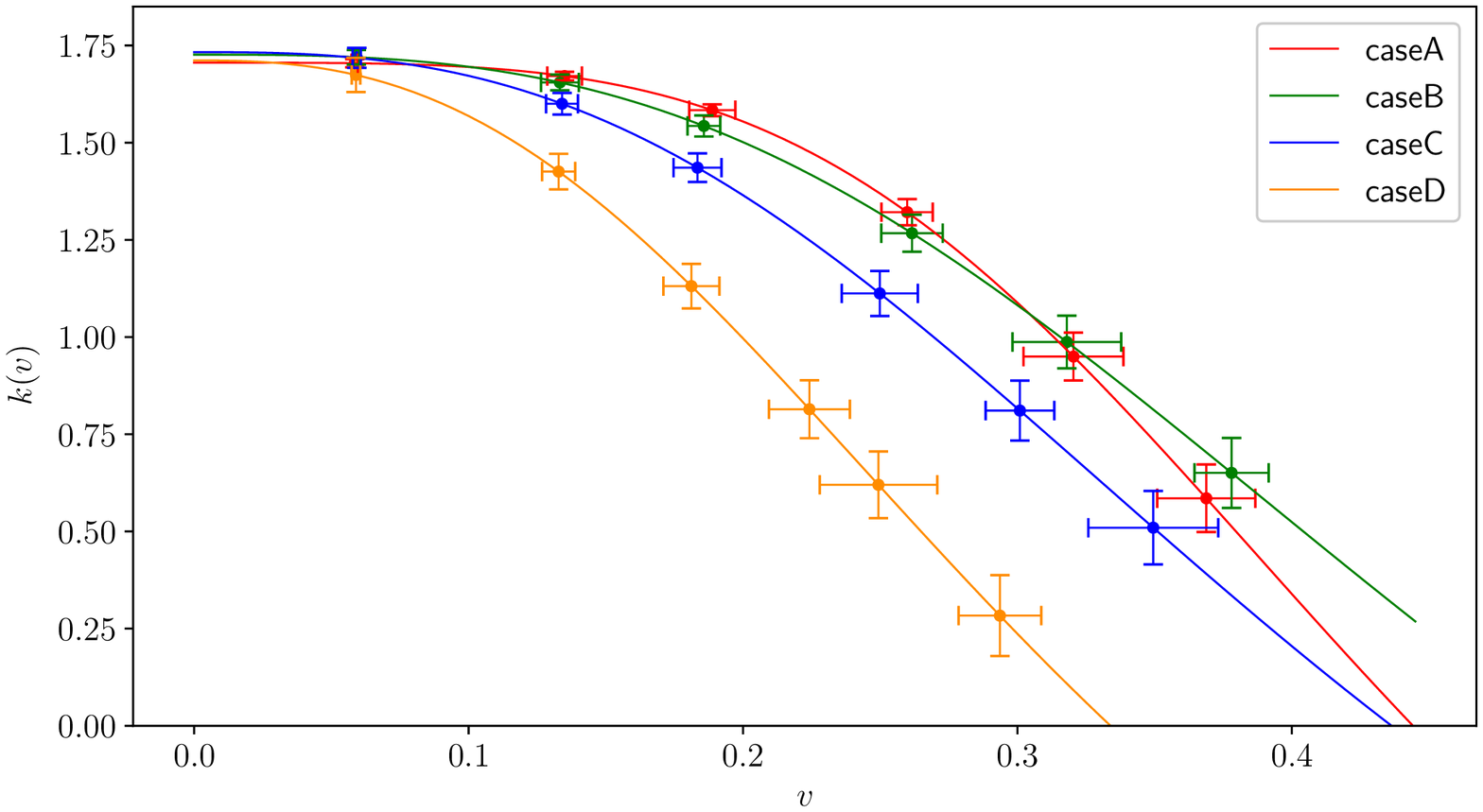}
\includegraphics[width=8.5cm]{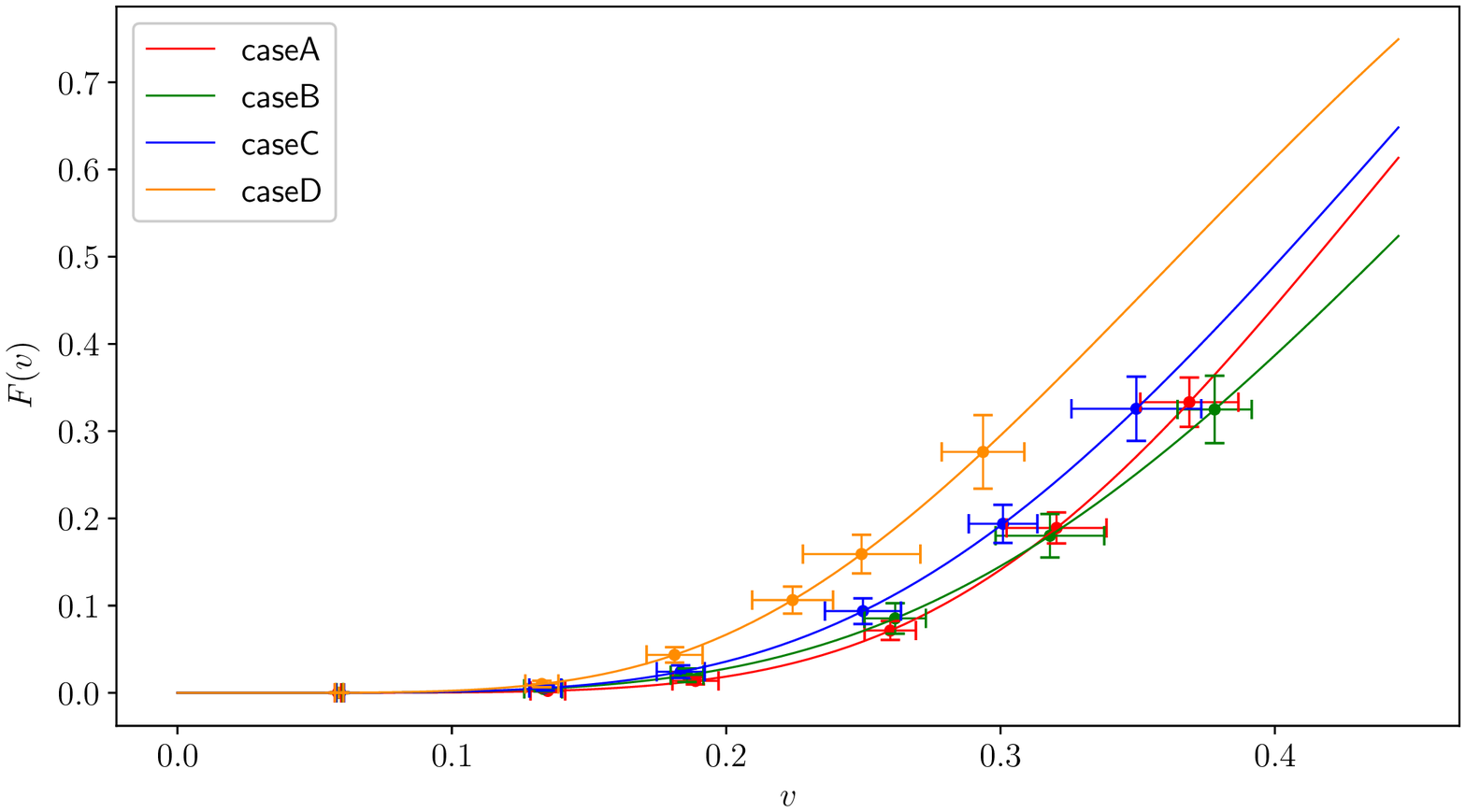}
    \caption{\label{fig:figure5}Comparing the VOS model prediction, using the best-fit parameters listed in \protect\ref{tab:param}, for the scaling parameters $\sigma / \rho \tau$ (top left), $v$ (top right), the curvature parameter $k(v)$ (bottom left) and the global energy losses parameter $F(v)$. The fits for Cases A, B, C and D are separately show in each panel.}
\end{figure*}

By solving the extended VOS equations for each case, we can also compare the model prediction with the measured quantities from the simulations. As can be seen in Fig. \ref{fig:figure5}, this comparison confirms that the modified model accurately describes the linear scaling regime of all cases.

\begin{figure*}
\includegraphics[width=8.5cm]{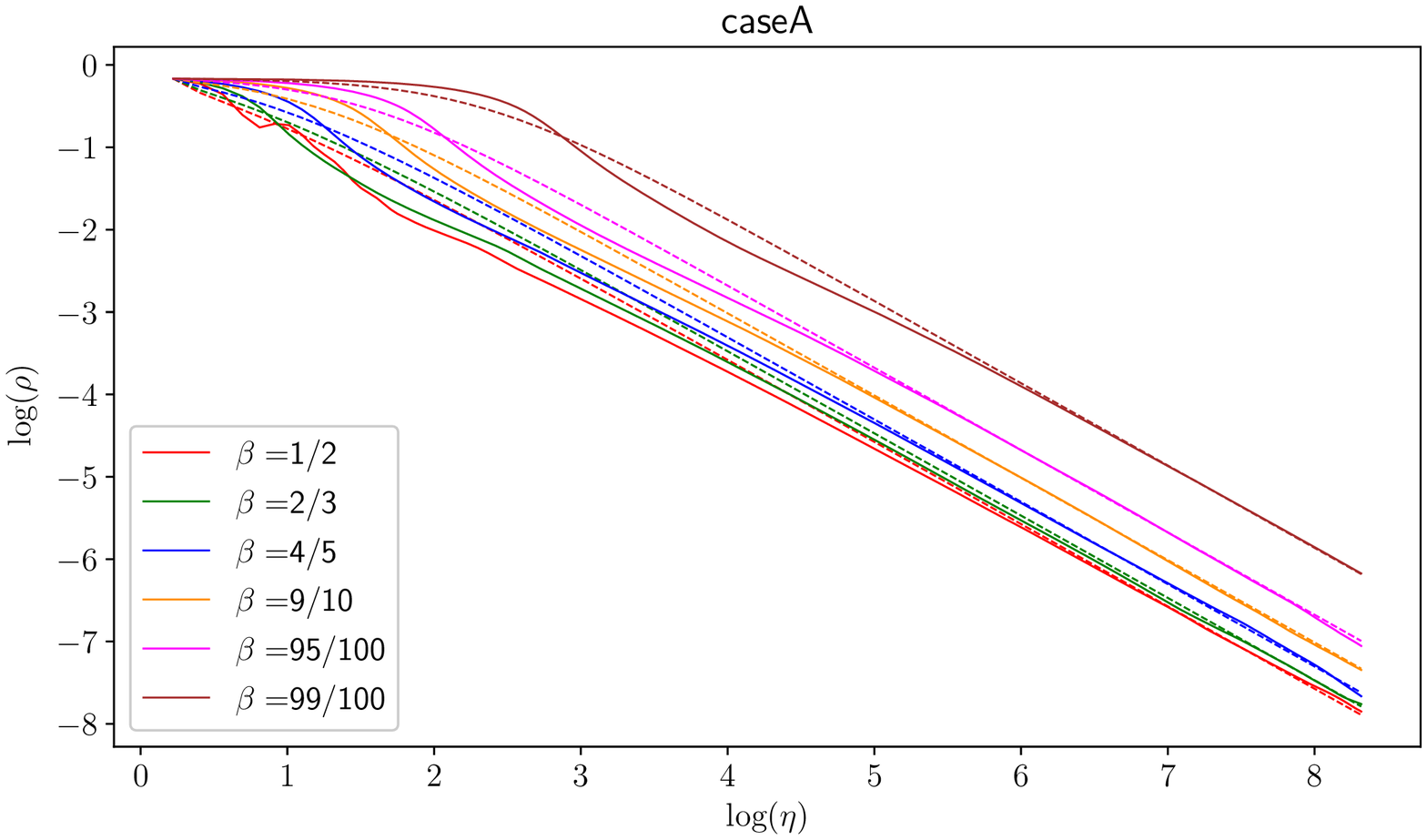}
\includegraphics[width=8.5cm]{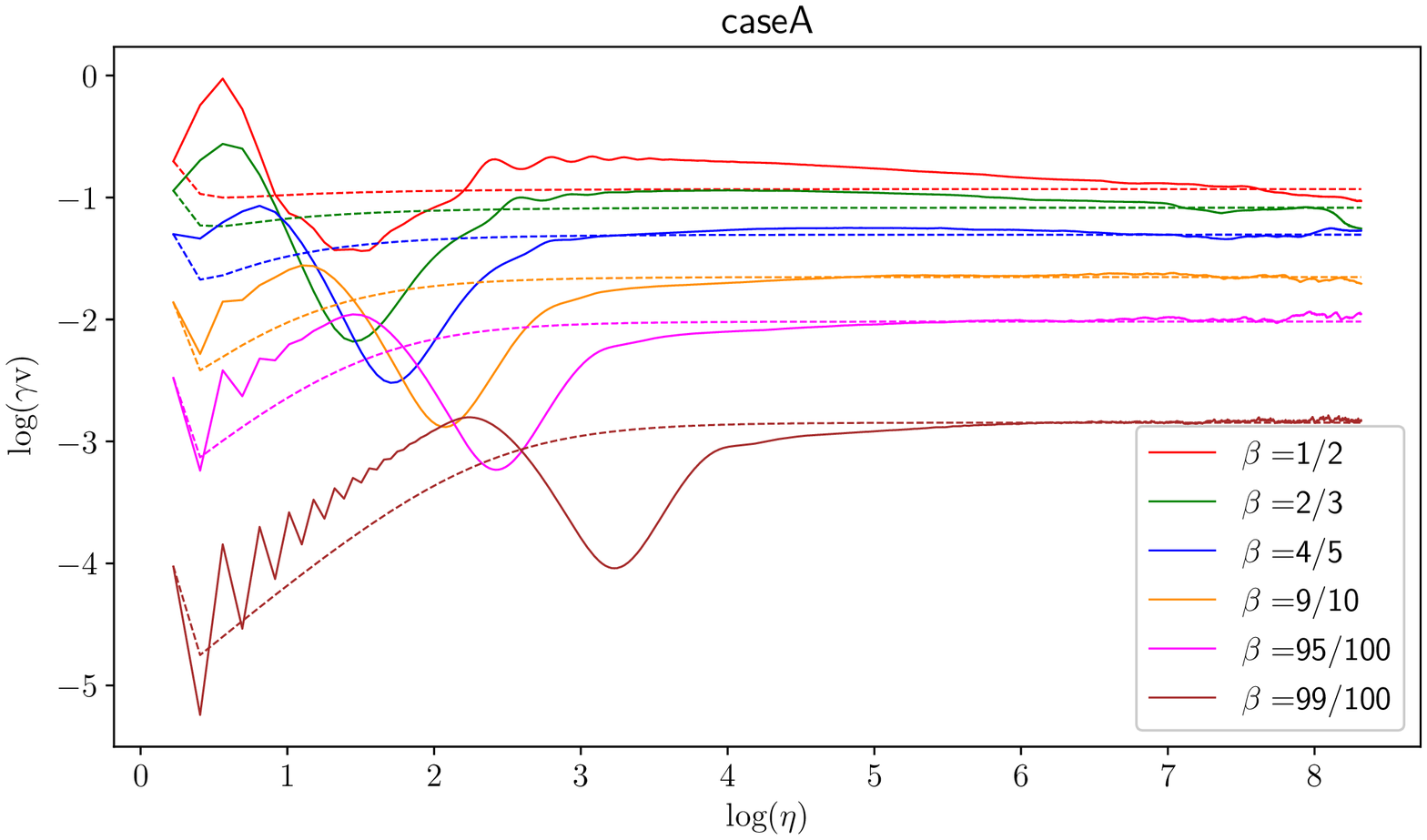}
\includegraphics[width=8.5cm]{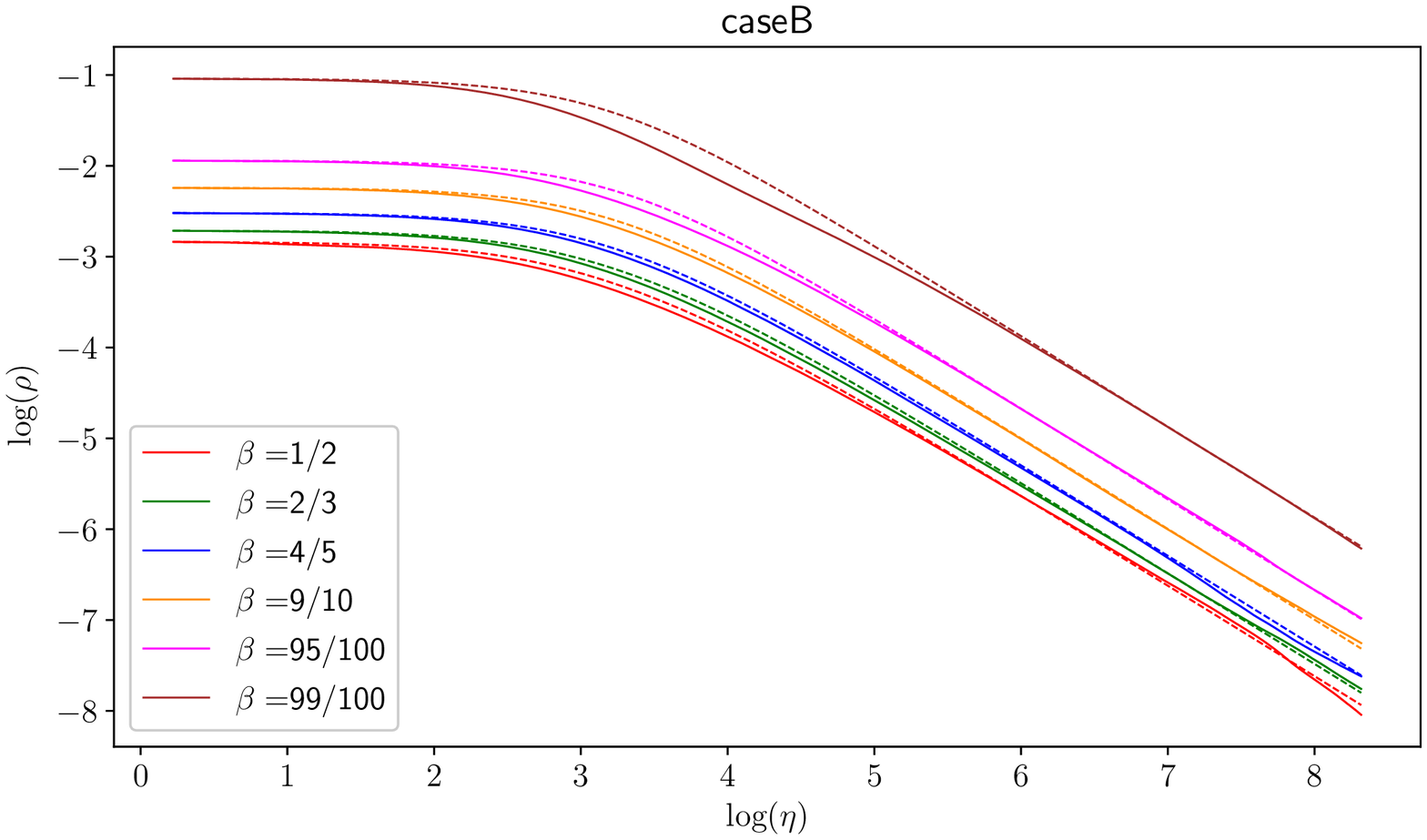}
\includegraphics[width=8.5cm]{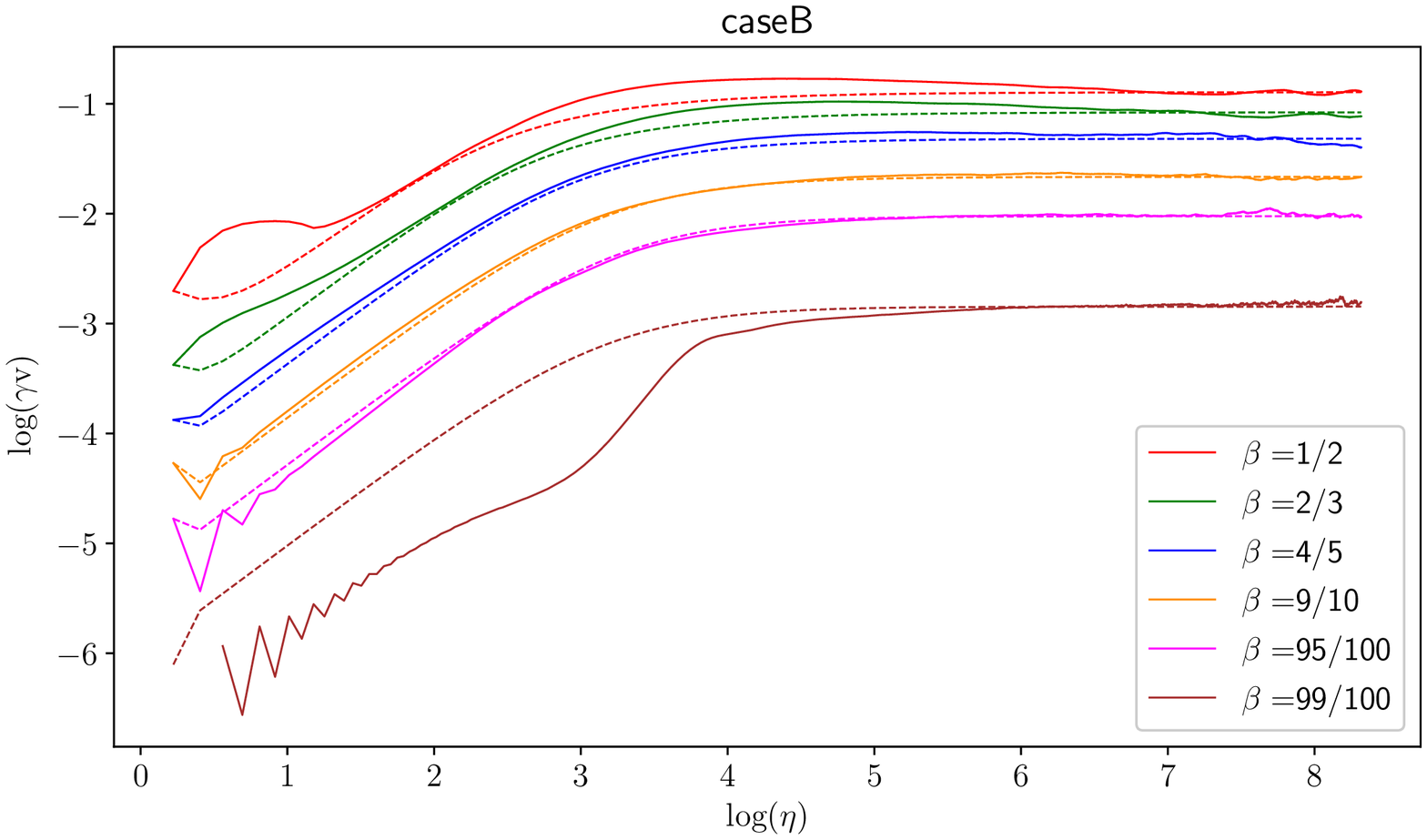}
\includegraphics[width=8.5cm]{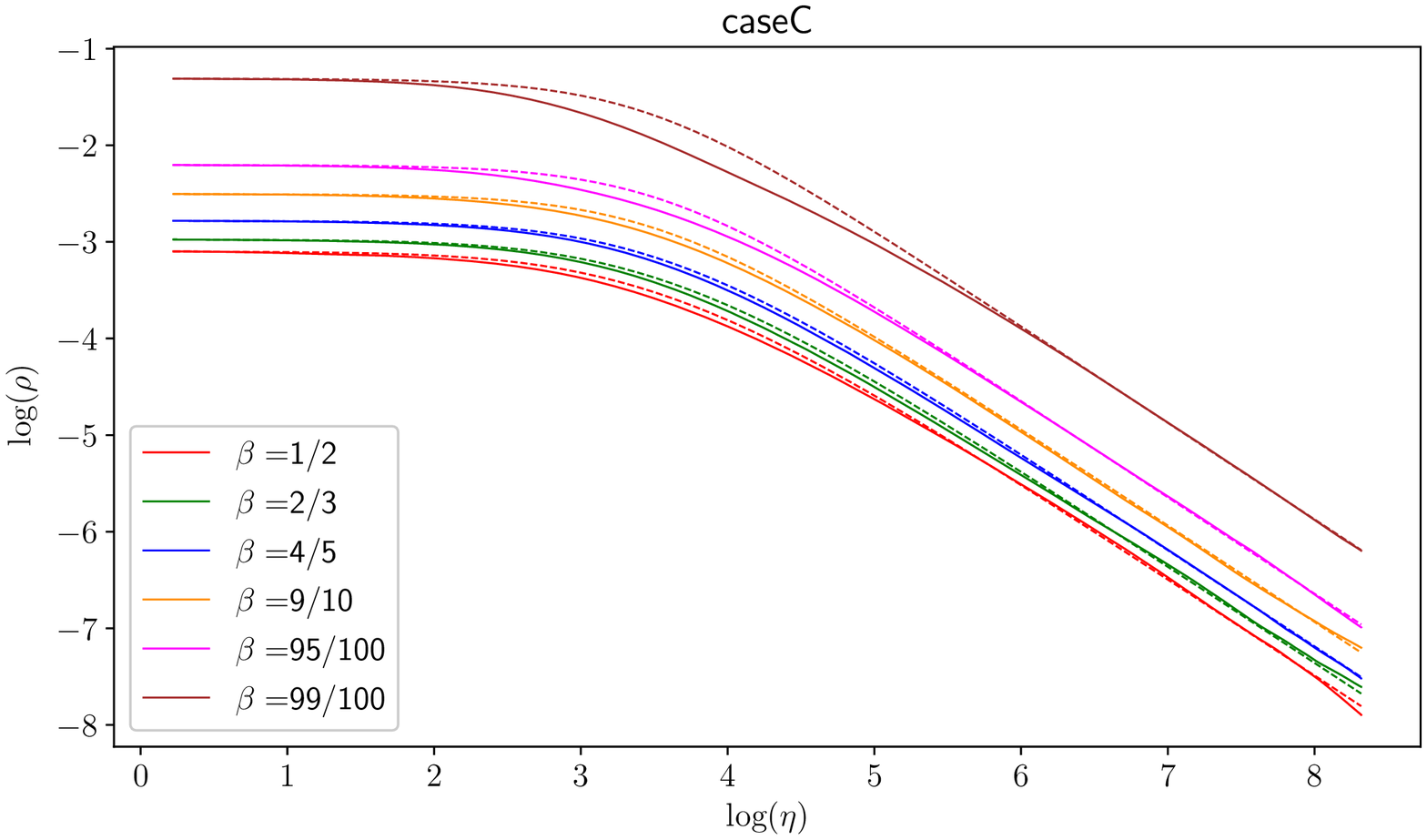}
\includegraphics[width=8.5cm]{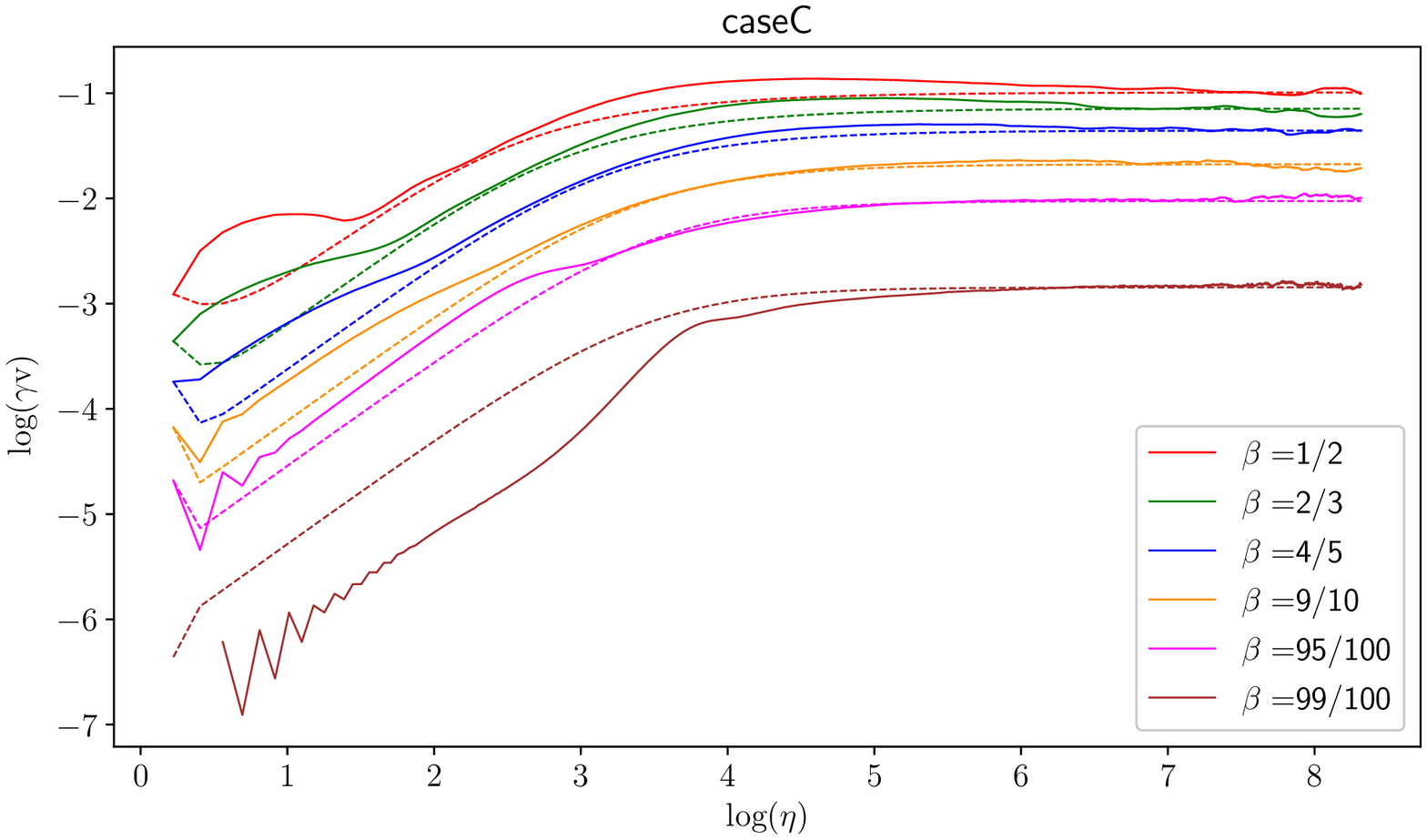}
\includegraphics[width=8.5cm]{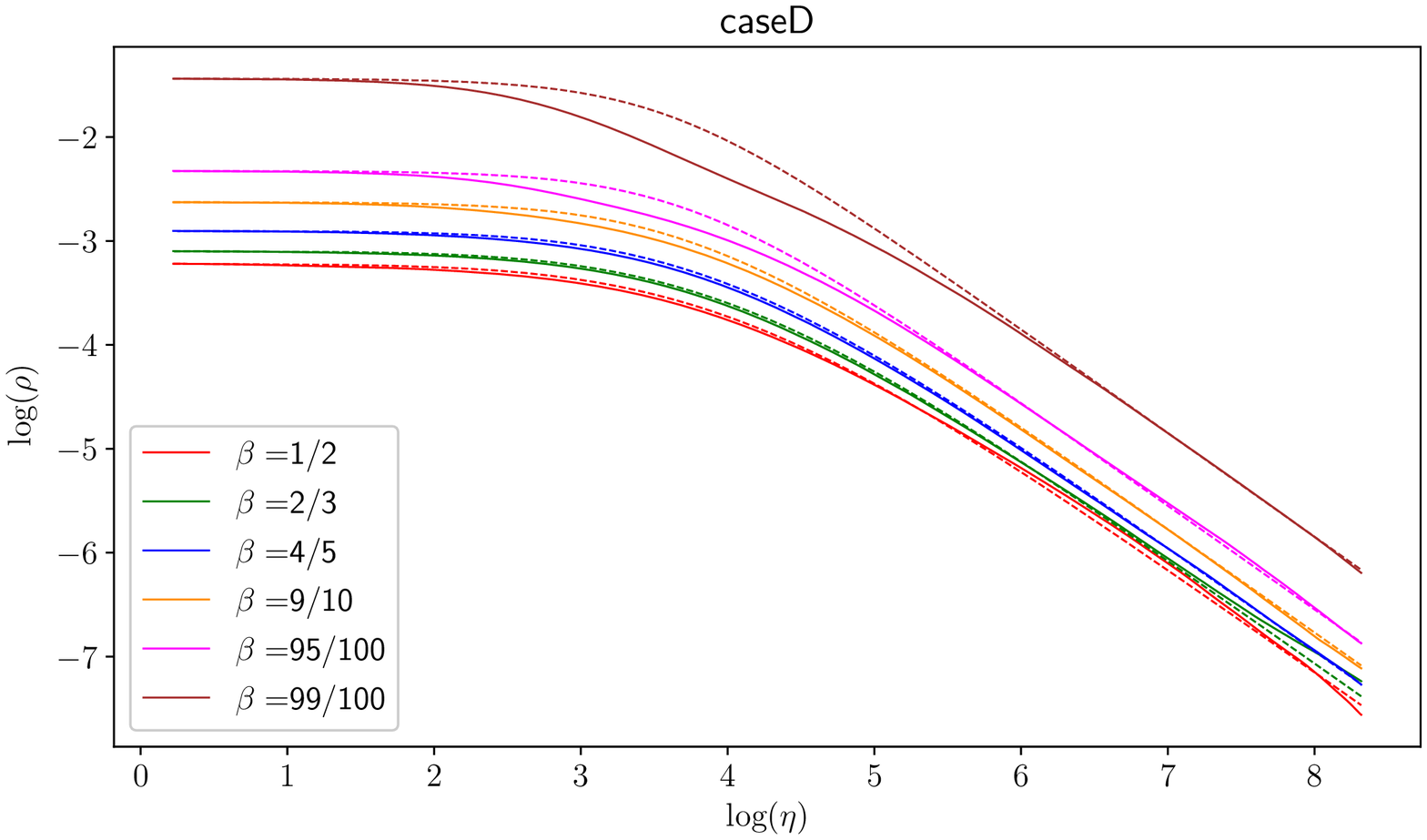}
\includegraphics[width=8.5cm]{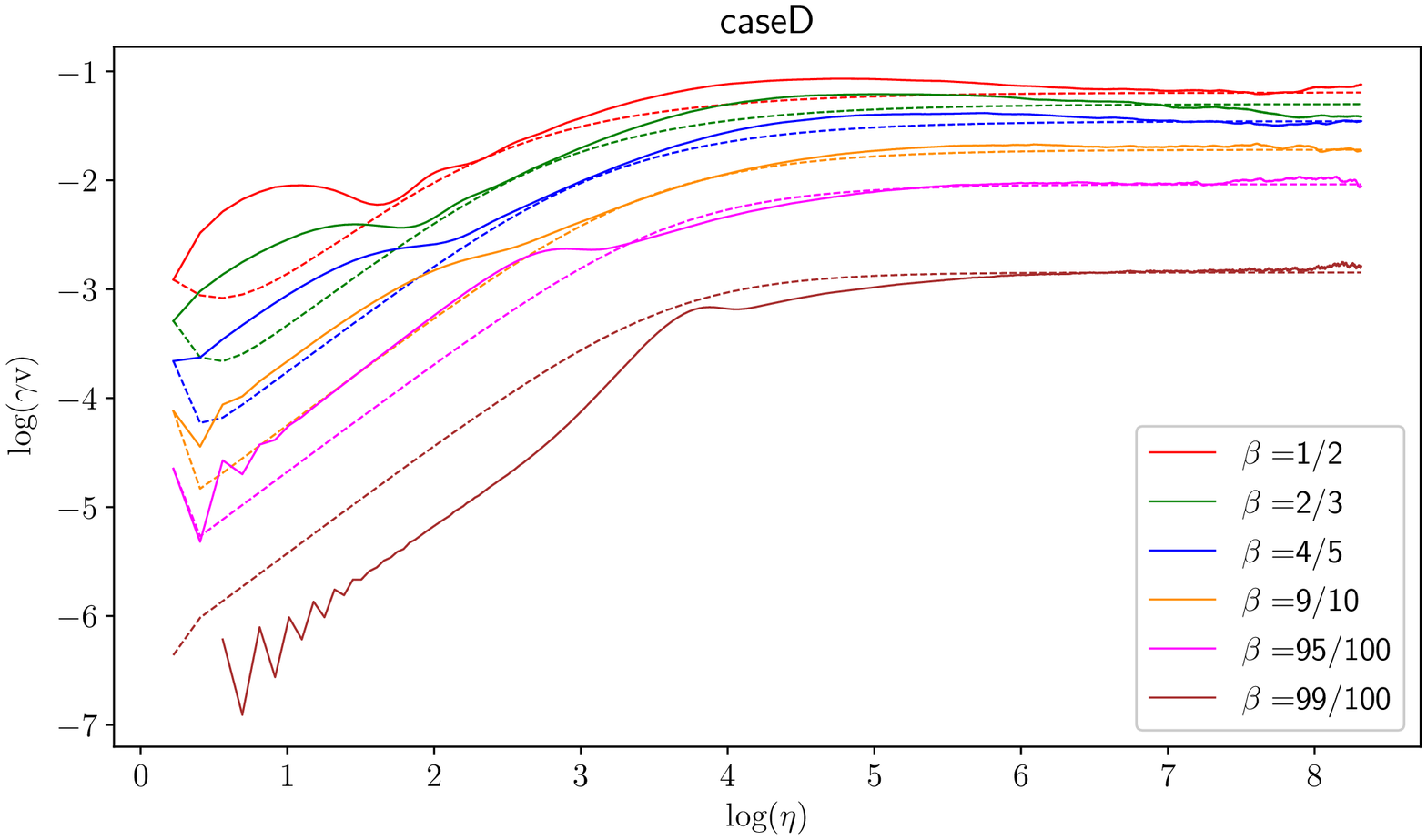}
\caption{\label{fig:figure6}Comparing the evolution of the density $\rho$ (left column) and the root mean squared velocity $\gamma v$ (right column) for simulations in the various cases and expansion rates. Each panel compares the results of the various expansion rates $\beta$ for one specific Case (A, B, C or D). In all panels the solid lines show the results of the simulations, while the dashed lines show the corresponding integration of the VOS, using the best-fit parameters of Table \protect\ref{tab:param}.}
\end{figure*}

\begin{figure*}
\includegraphics[width=8.5cm]{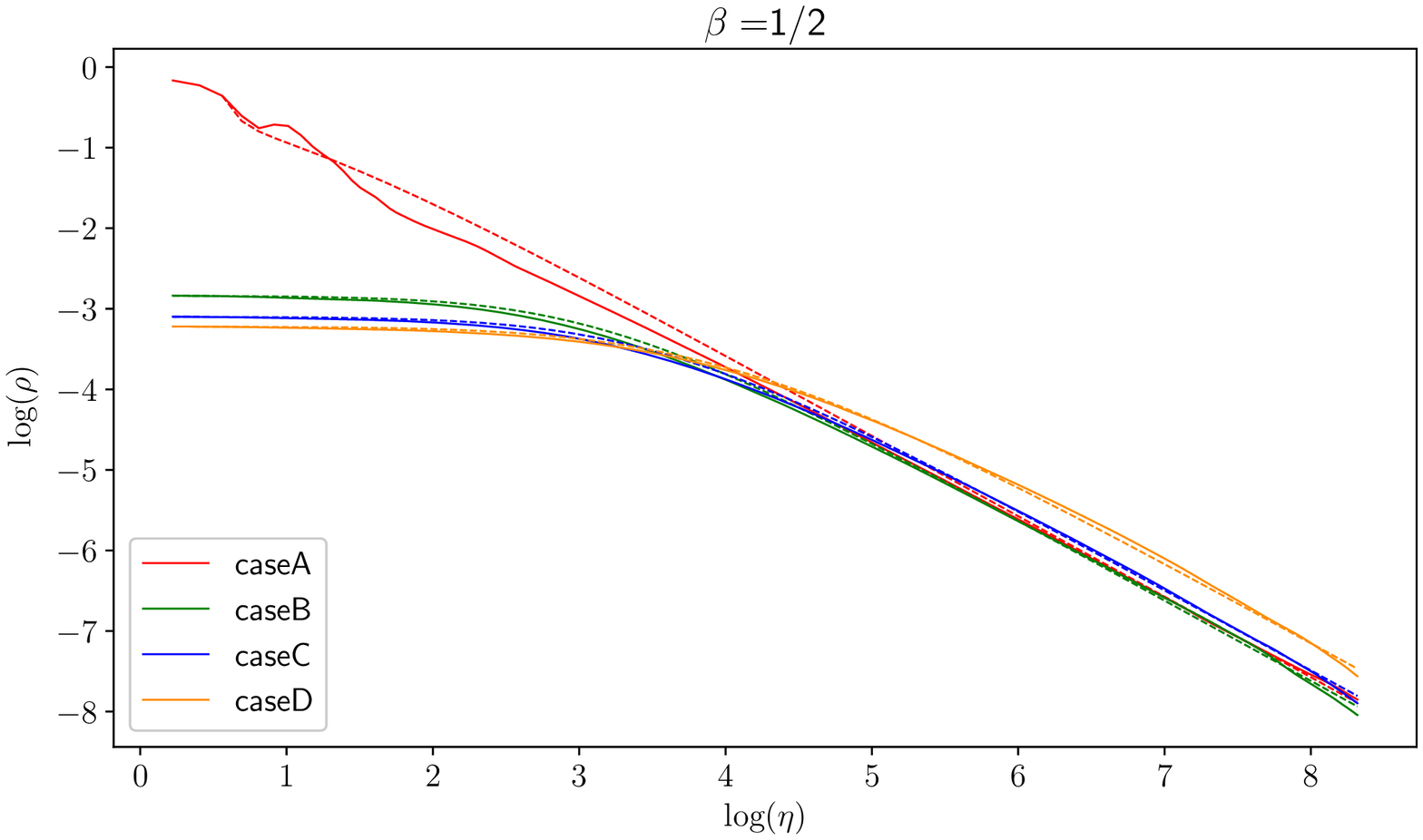}
\includegraphics[width=8.5cm]{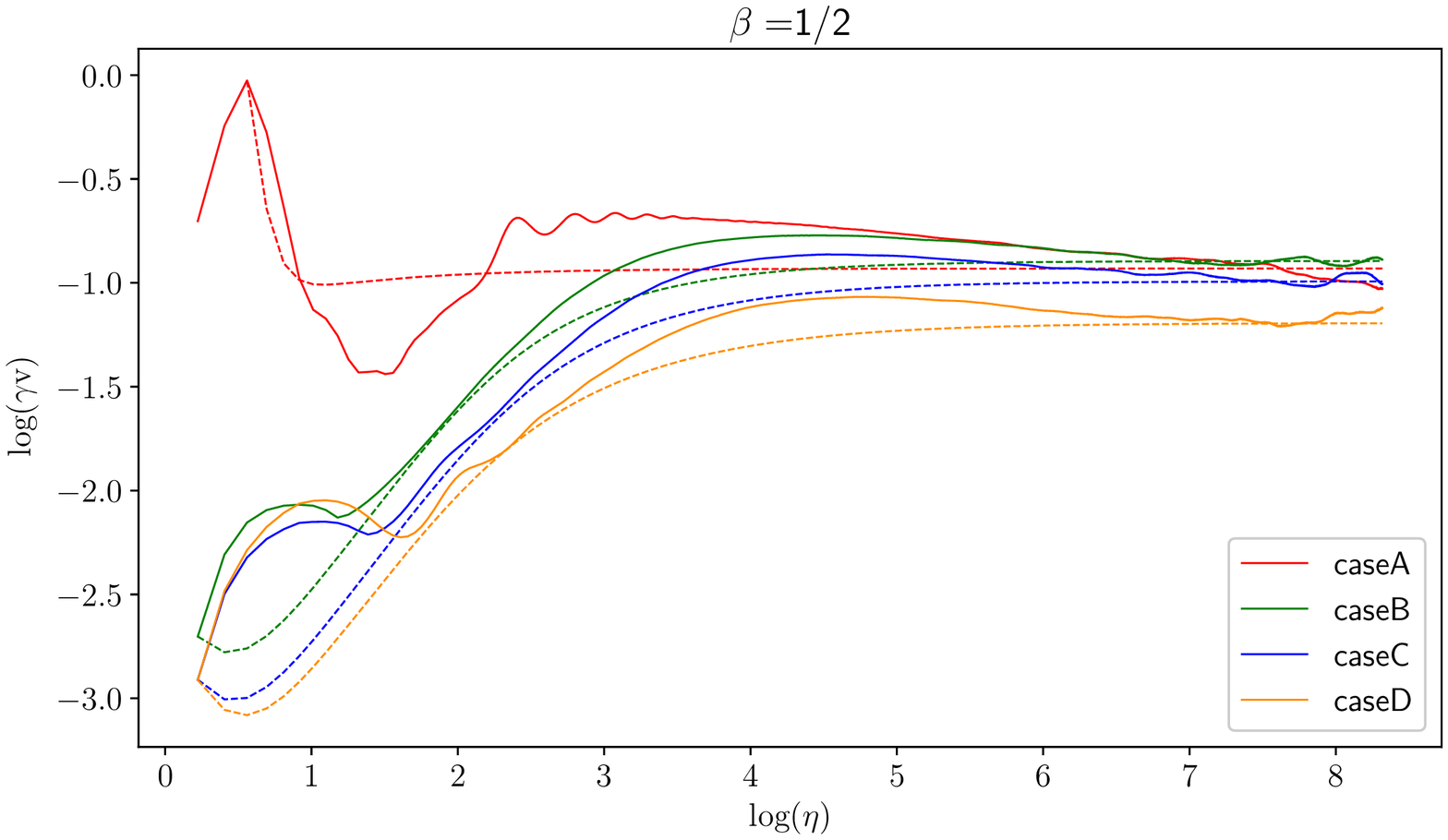}
\includegraphics[width=8.5cm]{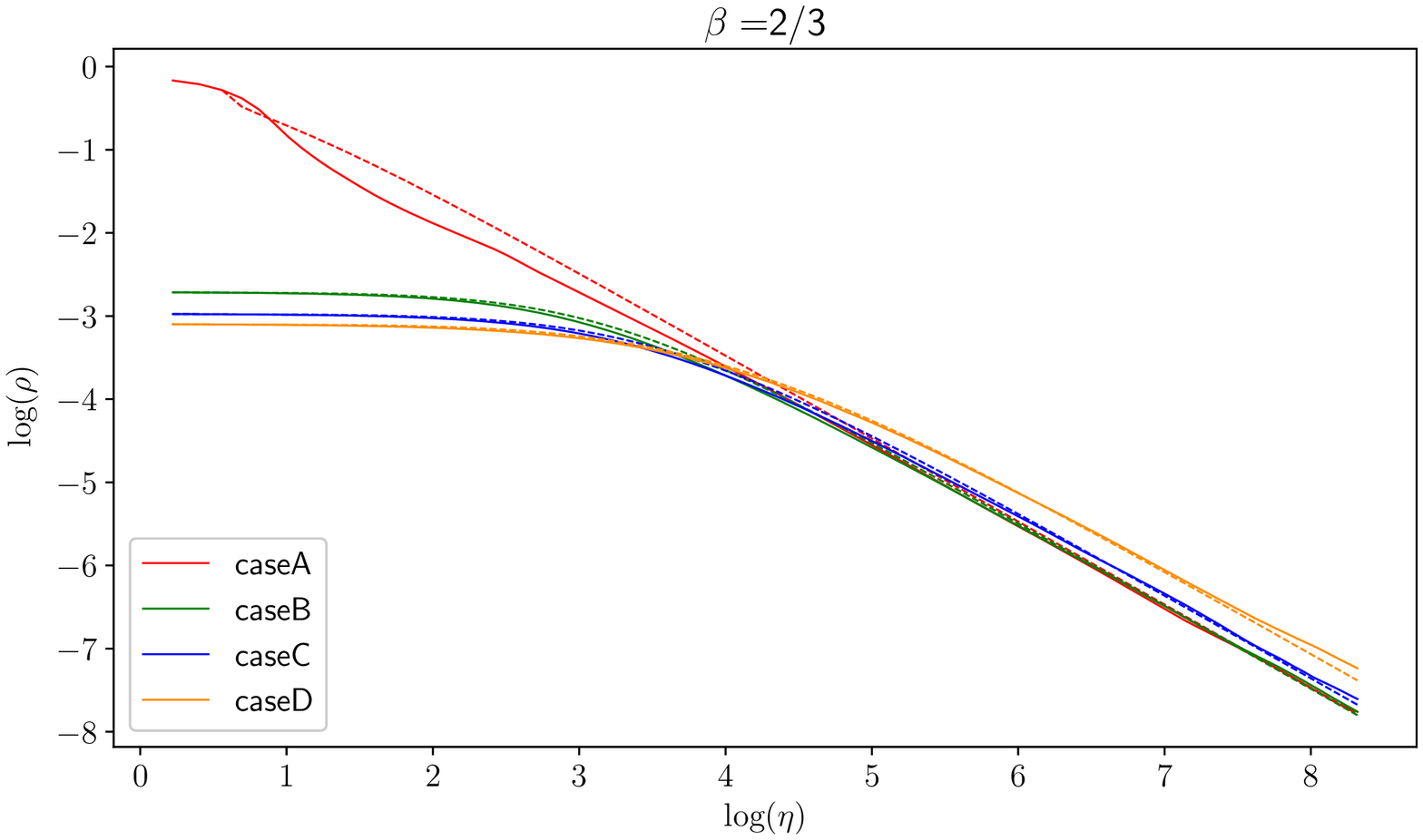}
\includegraphics[width=8.5cm]{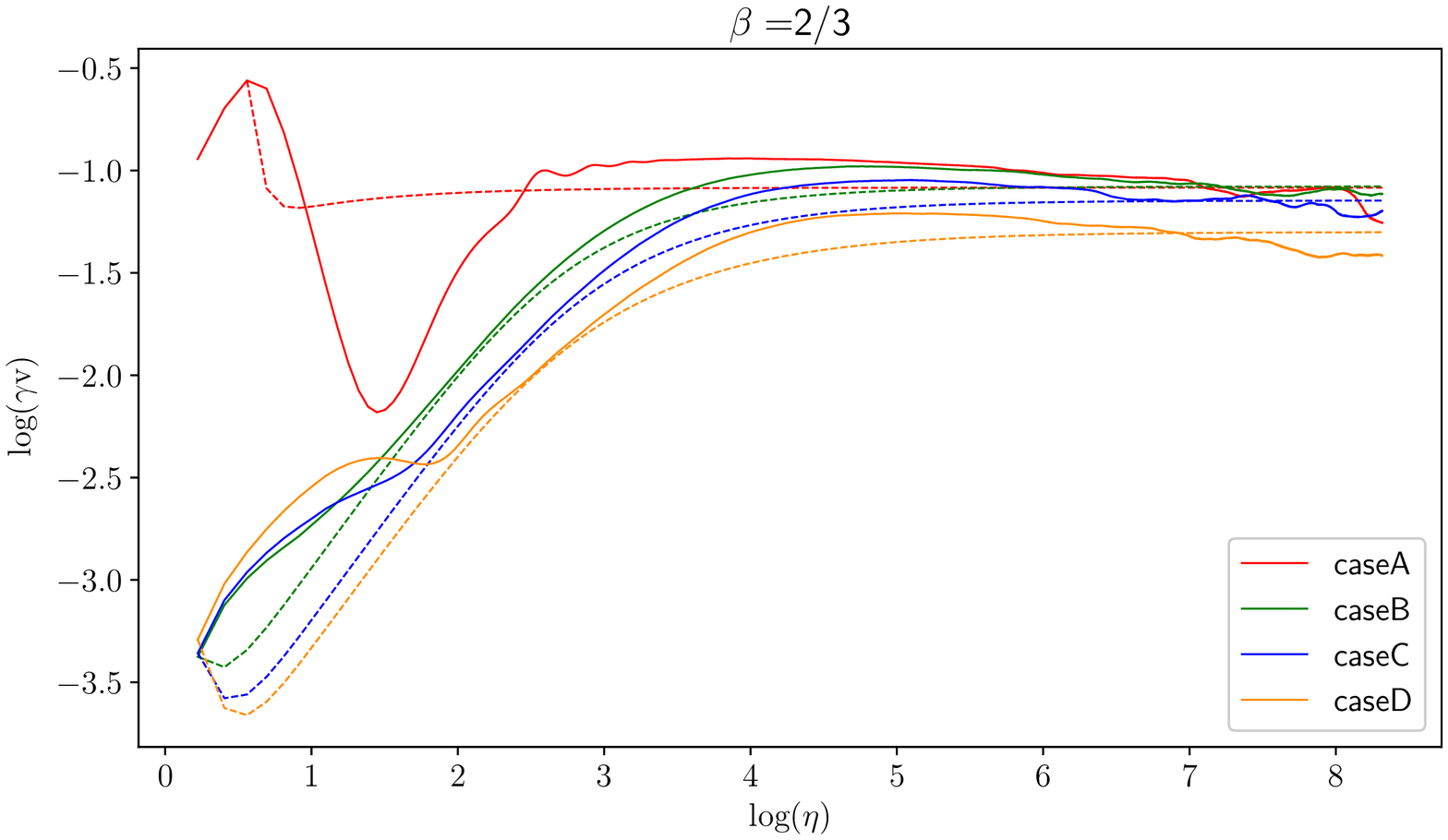}
\includegraphics[width=8.5cm]{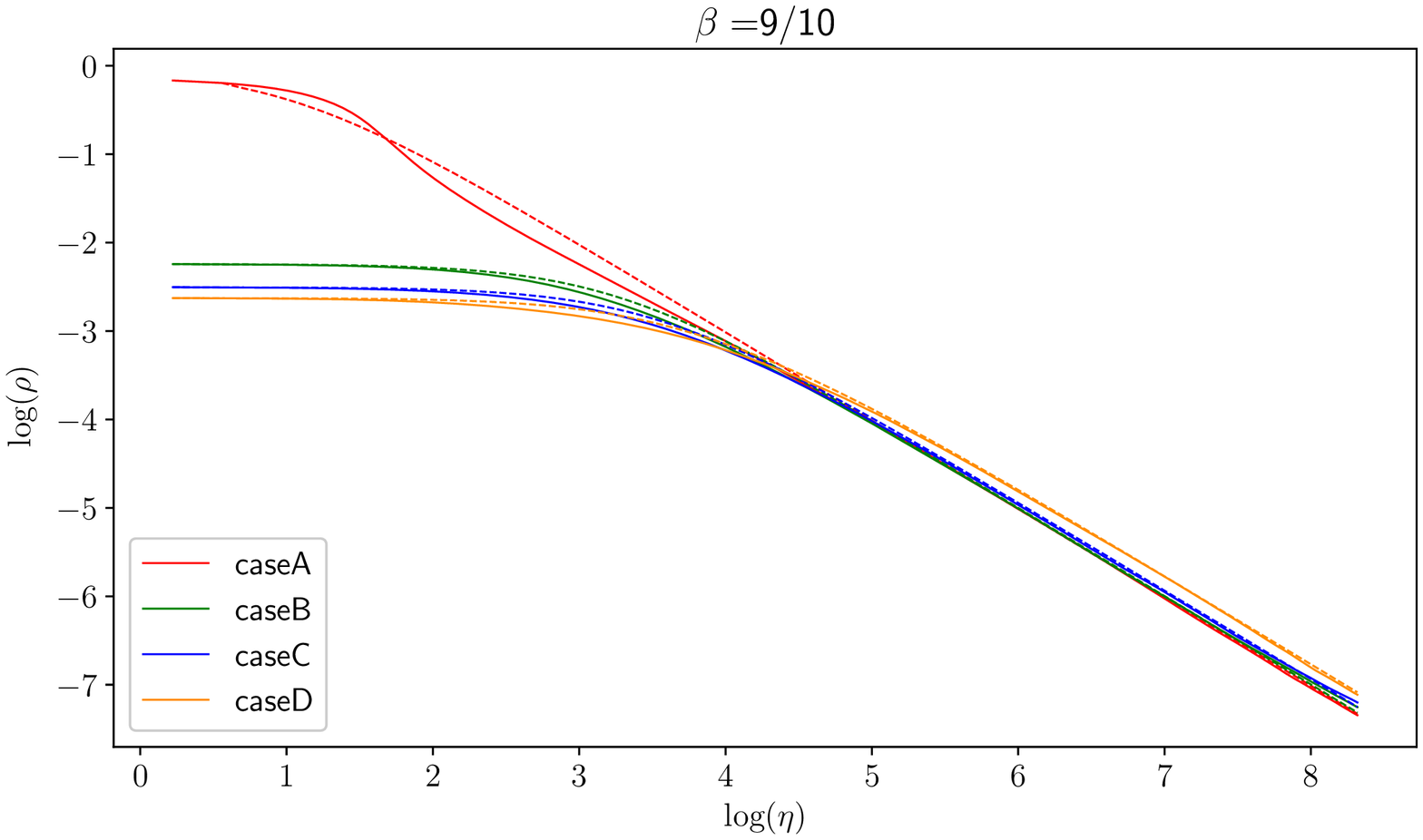}
\includegraphics[width=8.5cm]{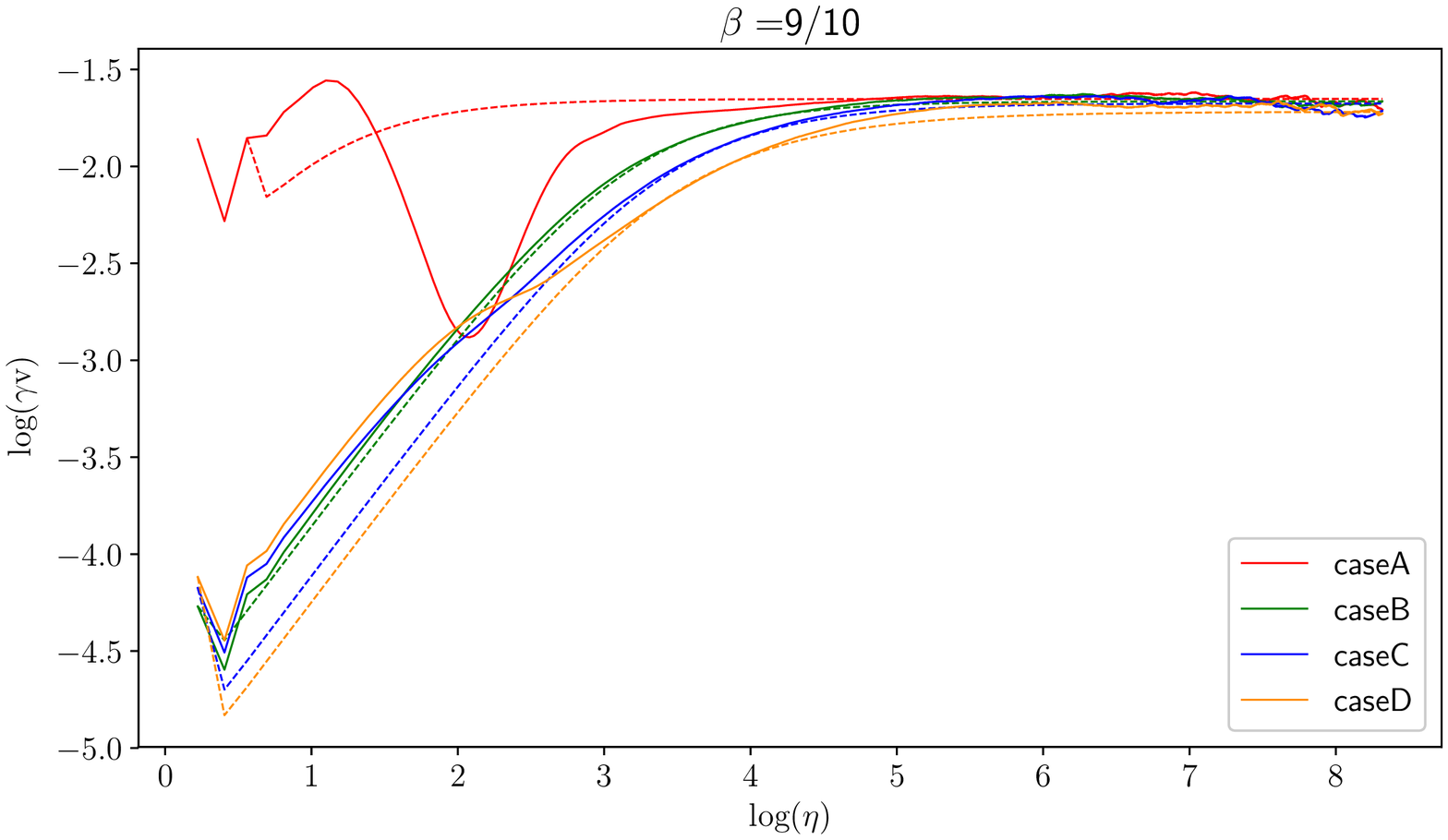}
\includegraphics[width=8.5cm]{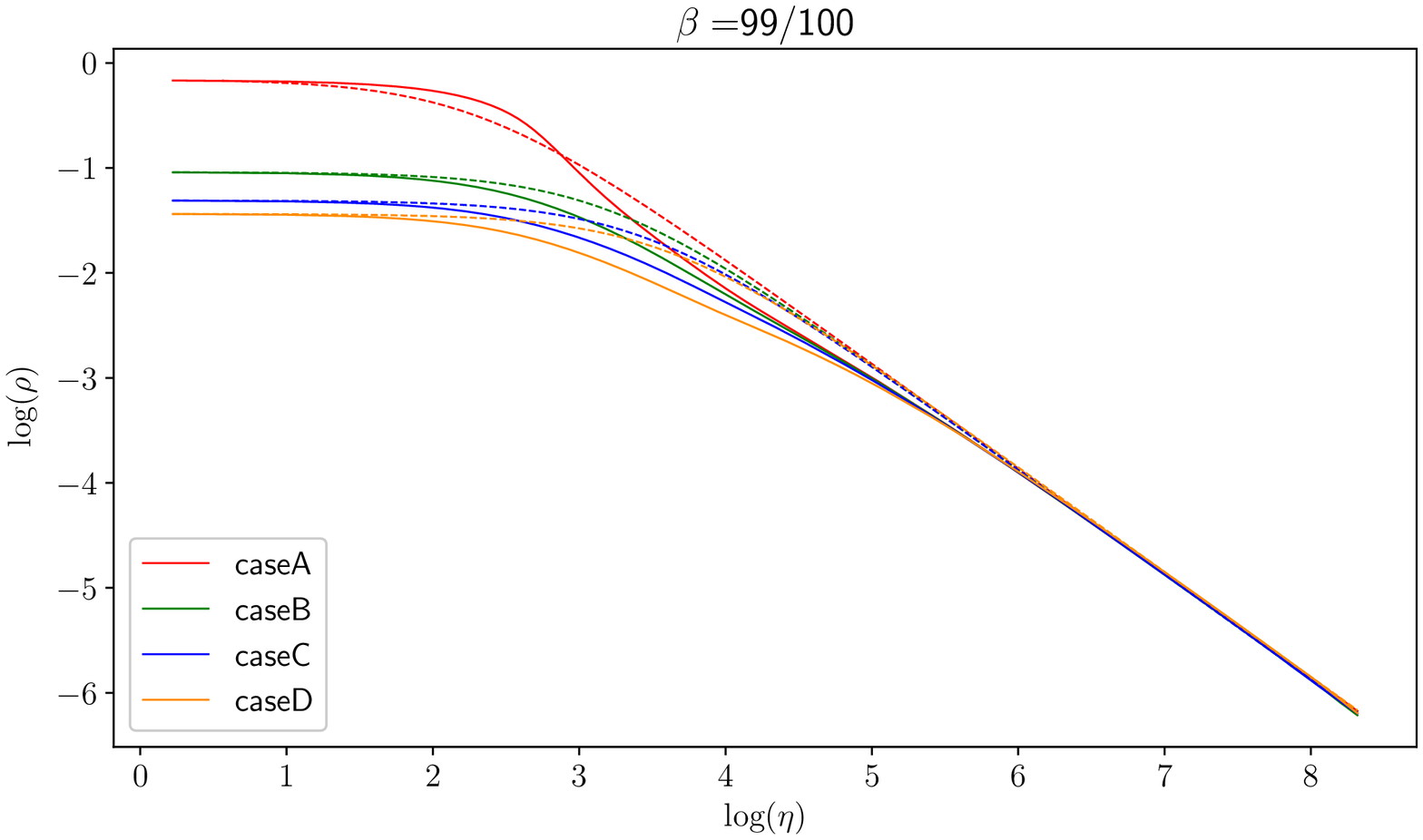}
\includegraphics[width=8.5cm]{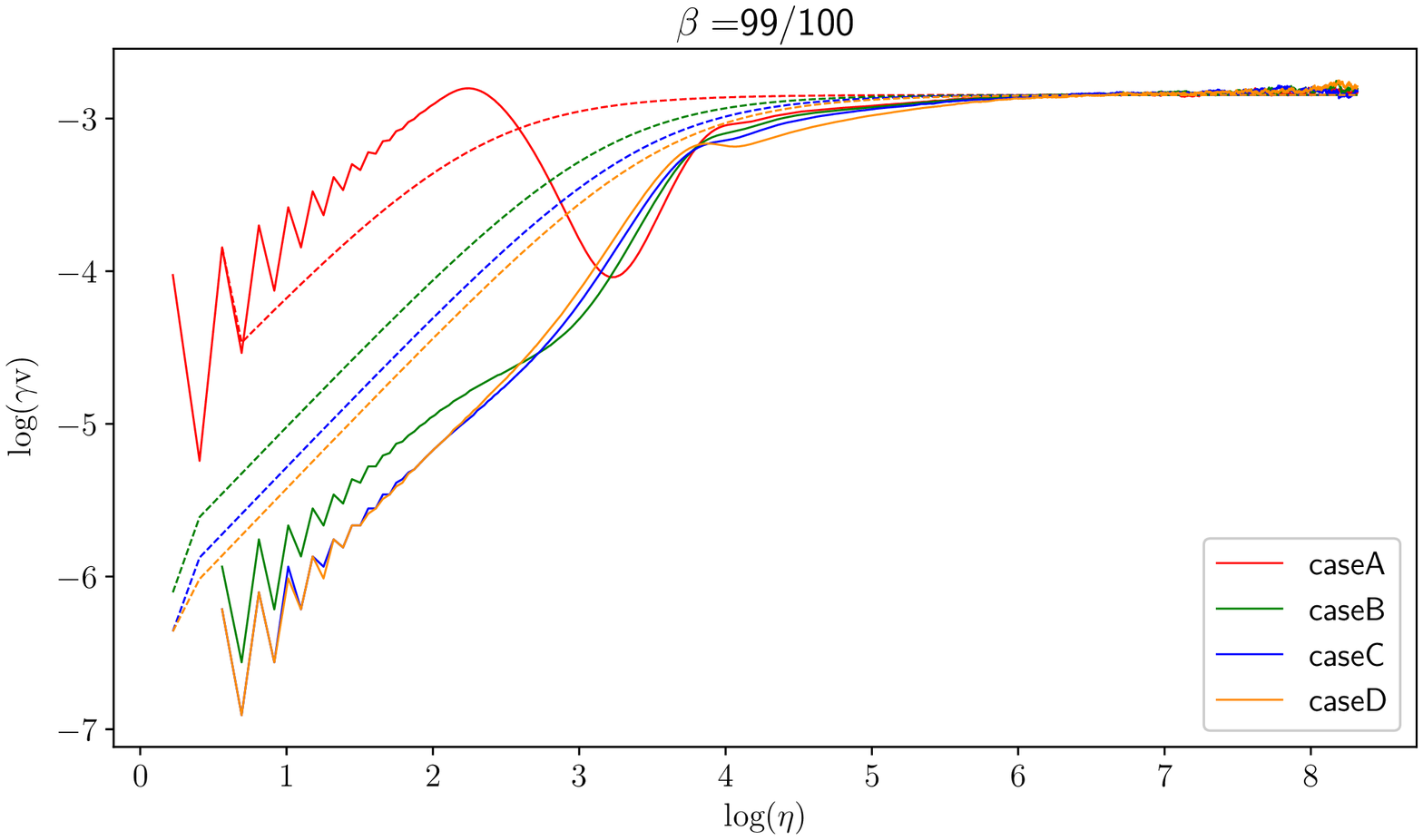}
\caption{\label{fig:figure7}As in Fig. \protect\ref{fig:figure6}, but each panel now compares the results of the four Cases (A, B, C and D) for a fixed expansion rate $\beta$; specifically the cases $\beta=1/2$ (radiation era), $\beta=2/3$ (matter era), $\beta=9/10$ and $\beta=99/100$ are shown.}
\end{figure*}

A further question is whether the VOS model can describe the approach towards scaling. Using the parameters of Table \ref{tab:param} we can numerically integrate the VOS equations for each expansion rate, using the first timestep of each simulation as initial condition. The results of this comparison are shown in Figs. \ref{fig:figure6} and \ref{fig:figure7}. We see that the model does this quite well for low expansion rates, and slightly less so for higher ones. This is due to the fact that in the latter case there is a different, transient scaling regime, first studied in \cite{Rybak2}.

\section{\label{concl}Conclusion}

In Paper I \cite{Previous}, we carried out a first comparative study of the evolution of various types of biased domain wall networks. In the present work we have extended Paper I in several ways, in particular quantifying some of the the earlier results and clarifying their physical interpretation.

For the cases of biased potential and biased initial conditions we have confirmed, by looking at the field distributions in the simulations, that the validity (or not) of the Gaussian approximation is crucial for the validity of the Hindmarsh decay law \cite{Hindmarsh} the key difference between the two cases. Specifically, we have confirmed that if one introduces a bias in the population fraction the Gaussian approximation is never realized, while if one introduces a bias in the canonical quartic potential the Gaussian approximation holds for some part of the evolution of the networks and the decay law is therefore obeyed.

For anisotropic walls we took advantage of our recently developed GPGPU walls evolution code to carry out a more systematic numerical exploration of the parameter space, in particular by simulating a broad range of expansion rates, enabling a quantitative comparison with the canonical VOS model for domain walls \cite{Rybak1}. Our results, based on $8192^2$ simulations, confirm that the model accurately predicts the linear scaling regime after isotropization, with model parameters consistent with those previously found from $4096^3$ simulations \cite{Rybak2}. Overall, our analysis provides a quantitative description of the cosmological evolution of these networks.

\begin{acknowledgments}
This work was done in the context of project PTDC/FIS/111725/2009 with additional support from grant UID/FIS/04434/2013. We gratefully acknowledge the support of NVIDIA Corporation with the donation of the Quadro P5000 GPU used for this research. J.R.C. is supported by an FCT fellowship (SFRH/BD/130445/2017). C.J.M. is supported by an FCT Research Professorship, contract reference IF/00064/2012, funded by FCT/MCTES (Portugal) and POPH/FSE (EC). J.R.C. would like to thank Ivan Rybak for the help given in the fits for the anisotropic walls section. 
\end{acknowledgments}

\end{document}